\documentclass[aps,prb,10pt,twocolumn,superscriptaddress,floatfix,amsmath,amssymb]{revtex4-1}

\usepackage{hyperref}
\usepackage[usenames,dvipsnames]{color}
\usepackage[english]{babel}
\usepackage{graphicx}
\usepackage{tabularx}

\begin{document} 
\title{Magnetic Susceptibility and Simulated Neutron Signal in the 2D Hubbard Model}

\author{J. P. F. LeBlanc}
\affiliation{Department of Physics and Physical Oceanography, Memorial
University of Newfoundland, St John's, Newfoundland and Labrador, Canada, A1B 3X7}
\affiliation{Department of Physics, University of Michigan, Ann Arbor, Michigan 48109, USA}
\author{Shaozhi Li}
\affiliation{Department of Physics, University of Michigan, Ann Arbor, Michigan 48109, USA}
\author{Xi Chen}
\affiliation{Department of Physics, University of Michigan, Ann Arbor, Michigan 48109, USA}
\affiliation{Center for Computational Quantum Physics, The Flatiron Institute, New York, New York, 10010, USA}
\author{Ryan Levy}
\affiliation{Department of Physics, University of Michigan, Ann Arbor, Michigan 48109, USA}
\affiliation{University of Illinois, Urbana-Champaign, Illinois 61820, USA}
\author{A. E. Antipov}
\affiliation{Department of Physics, University of Michigan, Ann Arbor, Michigan 48109, USA}
\affiliation{Microsoft Quantum, Microsoft Station Q, University of California, Santa Barbara, California 93106-6105 USA}
\author{Andrew J. Millis}
\affiliation{Department of Physics, Columbia University, New York, New York 10027, USA}
\affiliation{Center for Computational Quantum Physics, The Flatiron Institute, New York, New York, 10010, USA}
\author{Emanuel Gull}
\affiliation{Department of Physics, University of Michigan, Ann Arbor, Michigan 48109, USA}
\affiliation{Center for Computational Quantum Physics, The Flatiron Institute, New York, New York, 10010, USA}

\date{\today}
\begin{abstract}
We compute dynamic spin susceptibilities in the two-dimensional Hubbard model using the method of Dual Fermions and provide comparison to lattice Monte Carlo and cluster dynamical mean field theory. We examine the energy dispersion identified by peaks in ${\rm Im}\chi(\omega,q)$ which define spin modes and compare the exchange scale and magnon dispersion to neutron experiments on the parent La$_2$CuO$_4$ cuprate. We present the evolution of the spin excitations as a function of Hubbard interaction strengths and doping and explore the particle-hole asymmetry of the spin excitations.
We also study the correlation lengths and the spin excitation dispersion peak structure and find a `Y'-shaped dispersion similar to neutron results on doped HgBa$_2$CuO$_{4+\delta}$.
\end{abstract}
\maketitle

\section{Introduction}
The interplay between charge, spin and superconducting fluctuations in the cuprate high-temperature superconductors has been experimentally documented with nuclear magnetic resonance (NMR)\cite{yamada:2012}, inelastic neutron scattering (INS)\cite{Coldea01,Lipscombe09,lipscombe:2007,Vignolle07}, resonant ultrasound spectroscopy (RUS)\cite{shekhter:2013,leboeuf:2013}, thermal probes\cite{grissonnanche:2015,badoux:2016}, resonant inelastic X-ray scattering (RIXS)\cite{letacon:rixs,Braicovich10}, as well as Raman\cite{Devereaux07} and optical spectroscopies.\cite{Basov05} All together, they paint a  picture of a complex phase diagram comprised of many competing states where antiferromagnetic spin fluctuations play a crucial role.

Providing a theoretical description of even the normal high-temperature state of the cuprates has proven to be a formidable challenge.  Much has been understood on the level of single particle properties including the pseudogap in cluster DMFT studies of the Hubbard model\cite{Huscroft01,Parcollet04,Macridin06,Kyung06,Werner098site,Gull09_8site,Lin09,Liebsch09,sakai:2009,Lin10,Gull10_clustercompare,Sordi12,Lin12,Merino14,Gunnarsson15,Chen17} and its interplay with superconductivity.\cite{Lichtenstein00,Maier08,Civelli09,Gull12_energy,Gull13_super,Gull13_raman,Sordi13,Gull14_pairing,Gull15_qp,Chen15,Maier15,Maier18}
However, making a direct connection to many experimental measurements requires the knowledge of two-particle susceptibilities in addition to single-particle quantities. Among these, magnetic excitation effects have been a central focus.

The theoretical understanding of magnetic excitations remains an ongoing challenge.\cite{Chen15,gunnarsson:2015,Chen17} In the case of the insulating high-T$_c$ parent compounds, experimental data has primarily been fit using a linear spin-wave theory in the Heisenberg limit at zero temperature.\cite{Coldea01,delannoy:2009} This has allowed for a qualitative description of the excitations in terms of spin models and the determination of the strength of the exchange $J/t$.
However, spin models do not describe itinerant fermion systems, such as the doped compounds. In addition, the Heisenberg parameters determined via spin model fits lie outside the regime where spin-models are a valid low-energy approximation of the Hubbard model.
What is known to date about the  magnetic excitations of the Hubbard model and how they connect to experimental measurements on the cuprates \cite{Jia14, Zheng17, Huang:2017} comes from numerical studies of finite size systems. These methods accurately resolve short-range correlations but are limited in their temperature, doping, and momentum resolution and their ability to resolve long-ranged low-energy spin excitations.

In this paper we address this deficiency. We perform calculations on the Hubbard model in two dimensions using the technique of Dual Fermions (DF),\cite{Rubtsov08} which is an approximate extension of the non-perturbative dynamical mean field theory\cite{Georges96}  and is believed to be accurate at high temperature.\cite{Iskakov16} The method's primary advantage is that it can recover continuous momentum dependence, and therefore that it does not suffer from the finite system size effects that limit determinantal lattice quantum Monte Carlo (DQMC)  or cluster dynamical mean field theory results. In addition, the computational time is not substantially greater than that of single-site dynamical mean field theory, allowing us to probe the entire phase diagram with reasonable computational expense. We validate our results by comparing to cluster dynamical mean field theory in the dynamical cluster approximation (DCA) variant and DQMC in areas of parameter space that are accessible to those methods.
We show that for a certain interaction strength, the dispersion of spin-waves closely resembles that observed experimentally at high temperatures (300K) in La$_2$CuO$_4$.\cite{Coldea01}  We further examine the effects of both electron and hole doping, providing a complete picture of the particle-hole asymmetry of spin excitations in the model. 

The remainder of this paper proceeds as follows. In Section \ref{sec:methods} we discuss the DF method and provide comparisons to other methods.  Section \ref{sec:results} contains our main results from the DF method while section \ref{sec:discussion} concludes.

\section{Methods}\label{sec:methods}
We study the single orbital Hubbard model in two dimensions on a square lattice.\cite{scalapino:2007,LeBlanc15} The Hamiltonian is
\begin{align}
H = \sum_{k,\sigma} \left(\epsilon_{k} -\mu\right)c_{k\sigma}^\dagger c_{k\sigma}+U\sum_i n_{i\uparrow}n_{i\downarrow},
\label{eq:H}
\end{align}
where $\mu$ is the chemical potential, $k$ is momentum, $i$ labels sites in real-space, $U$ is the onsite Coulumb interaction, and the dispersion is given by
$\epsilon_k=-2t\left[\cos(k_x)+\cos(k_y)\right]-4t'\cos(k_x)\cos(k_y)$, in which $t$ and $t^{\prime}$ are the nearest and next-nearest neighbor hopping integrals.

\subsection{Dual Fermions}
We solve Eq.~\ref{eq:H} in the Dual Fermion approximation\cite{Rubtsov08} using the open source Dual Fermion code of Ref.~\onlinecite{Antipov15}. This method treats all local correlations in a non-perturbative manner and perturbatively includes non-local correlations. In our implementation, self-consistent ladder diagrams for the non-local (`dual') self energy are summed in the charge and spin channels.\cite{Hafermann09} This method is accurate at high temperature\cite{LeBlanc15} but uncontrolled, in the sense that contributions from higher order vertices and non-ladder diagrams are not included.\cite{Iskakov16} Results must therefore be carefully benchmarked against other techniques, both for single- and for two-particle properties.  To this end, in this section we present high temperature results from dynamical cluster approximation \cite{LeBlanc13,LeBlanc15,Iskakov16}  and from determinantal quantum Monte Carlo.\cite{BSS81}

\begin{figure}
\includegraphics[width=\linewidth]{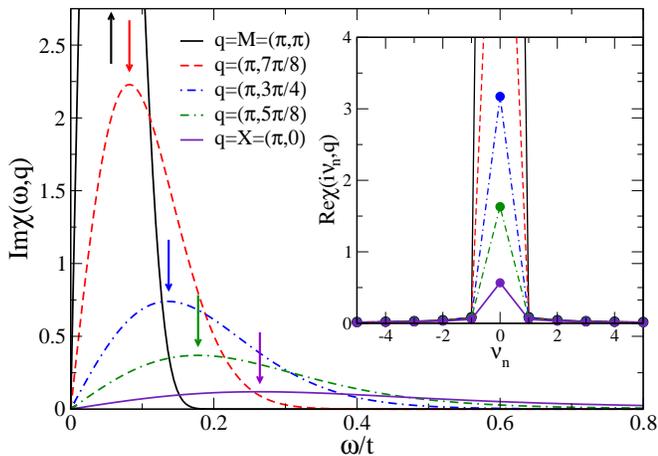}
\caption{\label{fig:realfreq} Dual Fermion results for the imaginary part of the Mabsubara axis susceptibility $\text{Im}\chi(\omega,q)$ as a function of real frequency $\omega$ for $U/t=8.0$ at $T/t=0.2$ with $t^{\prime}/t=-0.3$ at several $q$ vectors between the M and X points. Inset: Real part of the Matsubara axis susceptibility $\text{Re}\chi(i\nu_n, q)$ as a function of bosonic Matsubara frequency $\nu_n$ for the momenta $q$ indicated.
}\end{figure}

The quantity of interest is the spin susceptibility on the real frequency axis. First, we numerically compute spin susceptibilities on the Matsubara axis as a function of transfer momentum $q$ and bosonic Matsubara frequency $i\nu_n=2\pi i n T$. Representative simulation data obtained for a half-filled system at intermediate interaction $U/t=8$, temperature $T/t=0.2$, and next-nearest neighbor hopping strength $t'/t=-0.3$  are shown in the inset of Fig.~\ref{fig:realfreq}. Different curves denote susceptibilities at the momentum points indicated along a path from $X (\pi,0)$ to $M (\pi,\pi)$ in the Brillouin zone. 
Symmetry properties imply that only the real part of the susceptibility, $\text{Re}\chi(i\nu_n, q)$, is non-zero. 
Conversion of $i\nu_n$ to the real frequency, $\omega$, is needed to obtain the frequency- and momentum-dependent susceptibility $\text{Im}\chi(\omega,q)$.  This process relies on numerical methods of analytical continuation.  Those results can be seen in the main panel of Fig.~\ref{fig:realfreq}. $\text{Im}\chi(\omega,q)$ shows a single peak indicated by arrows. Its amplitude is largest at $q=(\pi,\pi)$ and quickly decays away from this point, while its peak position moves to higher frequencies. In the spirit of other works that examine spin structure factors and susceptibilities we define the frequency for this peak at each $q$-vector to be the spin-wave dispersion, $\omega_s(q)$.\cite{Jia14,Zheng17}

Analytic continuation of bosonic spectral functions, here based on the ALPS\cite{Gaenko17} open source maximum entropy code\cite{Levy2016} with a Gaussian default model, is required to obtain the spin-wave dispersion. The analytic continuation is numerically ill posed, in the sense that many bosonic spectral functions will yield the same bosonic Matsubara response within error bars. We believe continuation uncertainties to be larger than any finite size error or stochastic uncertainty. These uncertainties are independent of the approximation error of the self-consistent ladder DF method, which we analyze in detail below.

\subsection{Comparison to DCA}
\begin{figure}
\includegraphics[width=\linewidth]{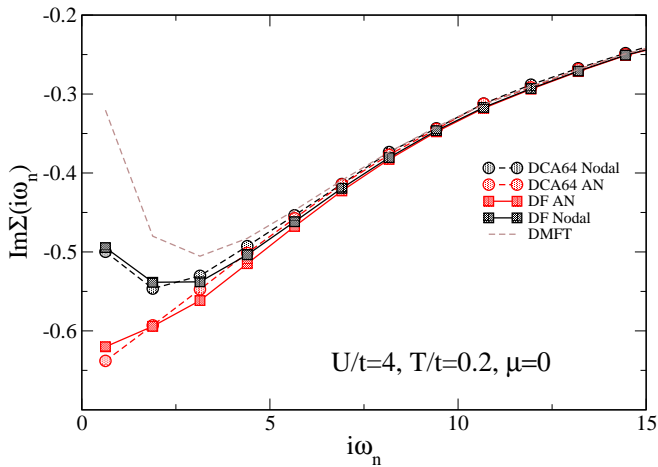}
\caption{\label{fig:dcadfU4} Data from 64-site DCA (dashed) and DF (solid) for the imaginary part of the self energy $\rm{Im}\Sigma(k,i\omega_n)$ at nodal (grey) and antinodal (red) points for $U/t=4$, $T/t=0.2$, $t'/t=0$, and half filling. }
\end{figure}

We first consider the accuracy of the DF method for single-particle quantities. 
 Fig.~\ref{fig:dcadfU4} compares results for the single-particle self-energy to  DCA calculations on an $8\times8$ cluster for the nodal ($k=(\pi/2,\pi/2)$) and antinodal ($k=(\pi,0)$) points.  The Dual Fermion result provides the majority of the momentum dependent contribution and is, at these temperatures and interaction strengths, comparable to DCA at a small fraction of the computational expense.

\begin{figure}
\includegraphics[width=\linewidth]{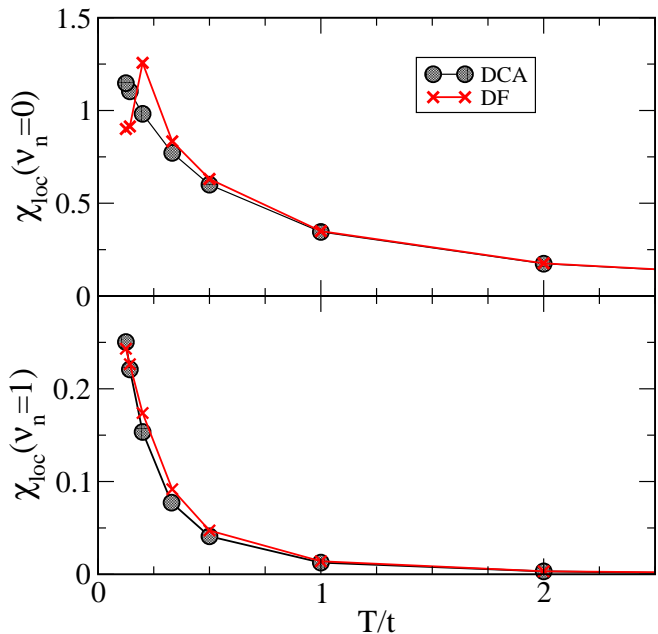}
\caption{\label{fig:dcadf} Data from 8-site DCA (black) and DF (red) for the local susceptibility at the lowest and first bosonic Matsubara frequencies, top and bottom frames respectively.  Data is for $U/t=6$, $t^{\prime}/t=-0.1$ at $\mu=0$. These calculations are for fixed $\mu$, not fixed density and some particle-hole asymmetry may be responsible for some variation in density with temperature between the two methods.}
\end{figure}
In Fig.~\ref{fig:dcadf} we compare the local spin susceptibility on the Matsubara axis for the lowest and first Matsubara frequencies obtained from Dual Fermions to those obtained in 8-site DCA clusters.\cite{Chen17}  Since DCA is limited in k-space resolution and computationally limited to small clusters when computing two-particle observables we compare only the local susceptibility $\chi_{loc}$.  We see agreement at high temperature where the spin susceptibility is expected to be unstructured and the DCA result is close to the dynamical mean field result upon which the Dual Fermion treatment is built.

As temperature is lowered, the lowest Matsubara frequency for DF is consistently larger than DCA until around $T/t=0.2$, below which it declines at $T/t=0.18$ and $0.19$.  At even lower $T/t$ the self-consistent ladder summation fails to converge. For this reason we consider $T=0.2t$ to be the lowest accessible temperature at $n=1$, $U/t=6$, and present results at this $T$ in section~\ref{sec:results}. 

\begin{figure}
\includegraphics[width=\linewidth]{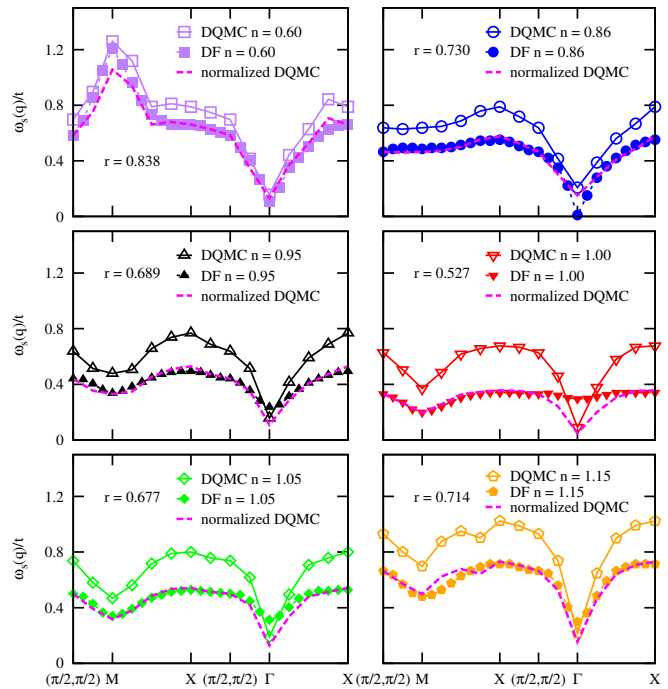}
\caption{\label{fig:figDQMC_n} DF (closed symbols) and DQMC (open symbols) results of the peak energy of the spin susceptibility, $\omega_s(q)/t$, extracted from $\text{Im}\chi(\omega,q)$ for high symmetry cuts in ($q_x$, $q_y$). Also shown (dotted lines) are DQMC results scaled by the factor $r$ given in each frame. $U/t=8$, $T/t=1/3$, and $t^\prime/t=-0.3$ illustrated for different doping $\langle n\rangle$.}
\end{figure}

\begin{figure}
\includegraphics[width=\linewidth]{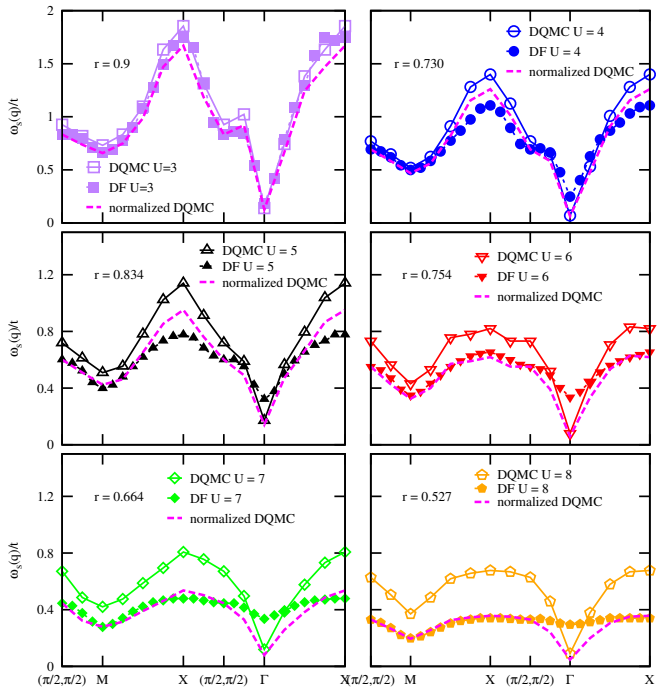}
\caption{\label{fig:figDQMC_U} DF and DQMC results of the peak energy of the spin susceptibility, $\omega_s(q)/t$, extracted from $\text{Im}\chi(\omega,q)$ for high symmetry cuts in ($q_x$, $q_y$). The parameters are $T/t=1/3$, and $t^\prime=-0.3$ for $\langle n\rangle=1$ illustrated for variation in interaction strength. The scaling of DQMC to DF data invokes a ratio, $r$, quoted in each frame. }
\end{figure}

\begin{figure}
\includegraphics[width=0.9\linewidth]{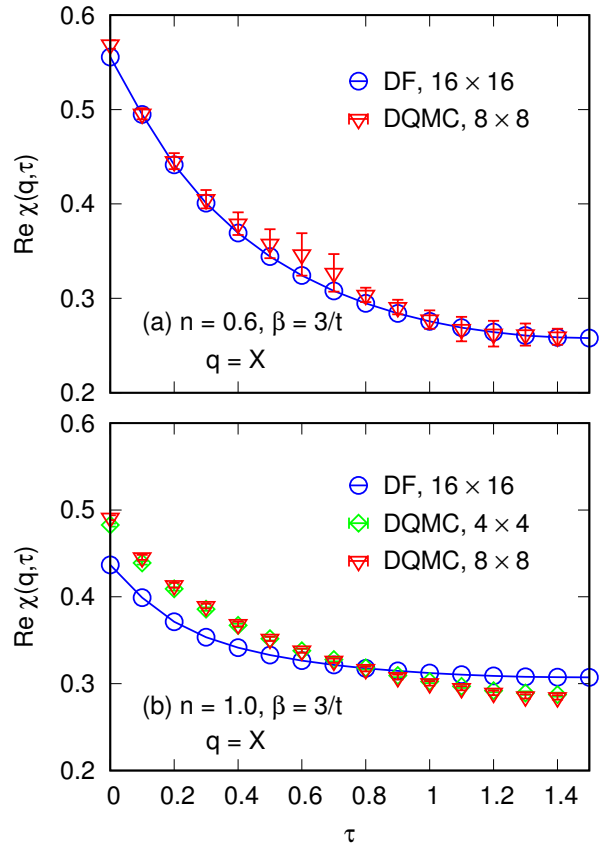}
\caption{\label{fig:figDQMC2} Spin susceptibilities $\text{Re}\chi(q,\tau)$ as a function of the imaginary time $\tau$ at $n=0.6$ (a) and $n=1.0$ (b).}
\end{figure}

\subsection{Comparison to DQMC}\label{sec:DQMC}
At high temperature, we can compare the dynamical spin susceptibility $\chi(q,\omega)$ on the real frequency axis obtained from DF to those obtained from DQMC calculations. In Fig.~\ref{fig:figDQMC_n} we plot the peak energy $\omega_s(q)$ of ${\rm Im} \chi(q,\omega)$ determined as in Fig.~\ref{fig:realfreq} as a function of momentum for various charge densities.  The momenta, along high symmetry cuts in the Brillouin zone, start from $(\pi/2,\pi/2)$ and move diagonally to the $M$-point at $(\pi,\pi)$, via $X=(\pi,0)$ and $(\pi/2,\pi/2)$ to $\Gamma$ and back to $X$. Here, the parameters are set as $U/t = 8$, $T/t = 1/3$, and $t^\prime = -0.3$. Dual Fermion calculations are evaluated on a $16\times 16$ grid (solid symbols), and DQMC calculations are performed on the $8\times 8$ cluster (open symbols). Both of these calculations are converged in their $k$-space discretization. At the lowest density ($n = 0.6$), Dual Fermion results are consistent with DQMC results. As we increase the density towards half-filling the spin excitation energy from Dual Fermion calculations are systematically lower than those from DQMC, but the momentum dependence is the same.
To illustrate this, we show in magenta on each panel the DQMC curve rescaled by the factor $r$ needed to match $\omega_s^{DF}$ at $q=(\pi/2,\pi/2)$ and quote the value of $r$ in each case.  After rescaling, we see that the overall structure of $\omega_s(q)$ is remarkably similar and that the approximate DF method is qualitatively correct within a single prefactor for each of the parameters examined, except near the $\Gamma$ point, as we comment on in section \ref{sec:discrepancy}. 

Continuing to increase charge density to $n=1.15$, the spin dispersion around the  $\Gamma$ point becomes identical, while the spin excitation at other momenta in Dual Fermion calculations has lower energy.
We perform a similar analysis in Fig.~\ref{fig:figDQMC_U} at fixed density, $n=1$, for variation in $U/t$.  We find that for small values of $U/t$ again the DQMC and DF are quantitatively in agreement while for other values they have a distinct energy scale with the DF result again systematically lower than the DQMC result but with the same momentum dependence.  The agreement in both highly doped and weak coupling cases but disagreement for larger interactions near half filling suggests that the method works best where the physics is close to that of the underlying DMFT approximation but may not be able to correctly represent the energy scales in the pseudogap and Mott insulating states.

To further analyze these conclusions, we plot the normalized magnetic susceptibility on the imaginary time axis, $\text{Re}\chi(q,\tau)$, for both calculations in Fig.~\ref{fig:figDQMC2}. These data do not suffer from continuation errors. Panel (a) shows $\text{Re}\chi(q,\tau)$ at $n=0.6$ and $q=X$. It is found that $\text{Re}\chi(q,\tau)$ for both calculations are the same, resulting in the same $\omega_s(q)/t$ in Fig.~\ref{fig:figDQMC_n} after analytic continuation. Panel~(b) shows $\text{Re}\chi(q,\tau)$ at $n=1.0$ and $q=X$ where the two methods deviate. To assess the finite lattice size effects in DQMC, we show results for two cluster sizes ($4\times4$ and $8\times8$).

Figs.~\ref{fig:figDQMC_n} - \ref{fig:figDQMC_U} together indicate that, at the high temperature tested here, the DF and DQMC are qualitatively the same. They differ quantitatively by a momentum independent prefactor that depends on both doping and interaction strength, $r\equiv r(U,\mu)$.  
As doped results from DCA or DQMC are difficult to obtain at lower temperature because of the fermion sign problem, we leave the temperature dependence of $r$ for future study.
We continue under the assumption that the $\omega_s(q)$ from  DF  is correct up to an overall doping-, temperature-, and interaction dependent prefactor.  This holds for all momenta except for the area near the $\Gamma$ point, which we address next.

\subsection{Discrepancy at $\Gamma$ point} \label{sec:discrepancy}
\begin{figure}
\includegraphics[width=\linewidth]{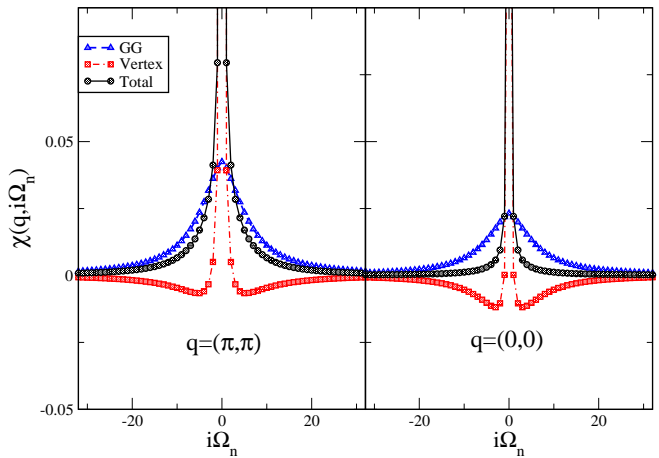}
\caption{\label{fig:bubble-vertex} $\rm{Re}\chi(i\Omega_n,q)$ for $q=(\pi,\pi)$ (left) and $q=(0,0)$ (right) at $U/t=8$, $T/t=0.2$, $t'/t=-0.3$, and $n=1$.  The total susceptibility (black) is decomposed into contributions from the dressed polarization bubble, $GG$, shown in blue and the vertex contributions shown in red.}
\end{figure}

\begin{figure}
\includegraphics[width=\linewidth]{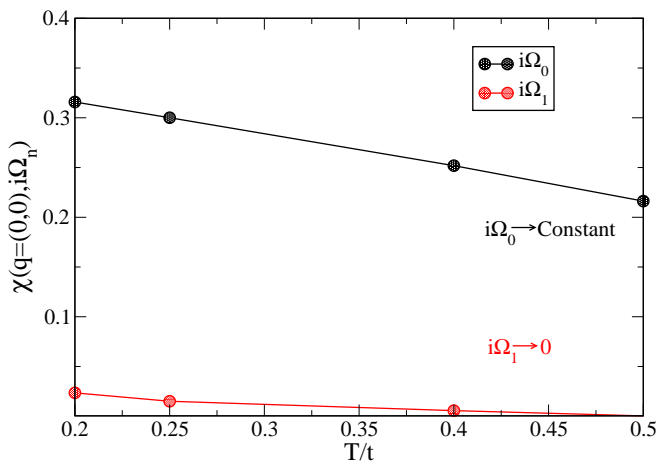}
\caption{\label{fig:chi_tvary} $\chi$ for the zeroth and first bosonic Matsubara frequencies as a function of $T/t$. The other parameters are the same as in Fig. \ref{fig:bubble-vertex}}
\end{figure}

Figure \ref{fig:figDQMC_n} shows that $\omega_s(q=\Gamma)$ at $n=1$ does not touch to zero in DF calculations.
However, since the spin operator commutes with the Hamiltonian, the spin susceptibility on the real frequency at the $\Gamma$ point should be uniformly zero, {\it i.e.} $\chi(q=\Gamma, \omega)=0$ for all values of $\omega$. On the Matsubara axis, this condition enforcing a spin excitation at zero frequency is that $\chi(q=\Gamma, i\Omega_0)$ is a constant while $\chi(q=\Gamma, i\Omega_n)=0$ as $n\ne 0$. 
To understand why DF results violate this condition, we plot spin susceptilibilty $\chi(q,\Omega_n)$ and its dressed bubble (GG) and vertex terms in Fig. \ref{fig:bubble-vertex}. 
The spin susceptibility equals the sum of the GG and the vertex terms. The right panel of Fig. \ref{fig:bubble-vertex} shows that vertex contribution does not exactly cancel the bubble contribution at $i\Omega_n$ for $n\ne 0$. Other methods of restoration exist.\cite{rubtsov:2012,vanloon:2014}
 The magnitude of this violation is temperature dependent. As temperature increases, the value of the first Matsubara frequency decreases, see Fig. \ref{fig:chi_tvary}, and at high temperature ($T/t=0.5$) becomes essentially uniformly zero, satisfying the expectation that $\chi(q=\Gamma, \omega)=0$ for $\omega=0$.

The violation of this cancellation is induced by the approximate nature of the Dual Fermion procedure, which is based on a vertex function expansion of an auxiliary Anderson impurity coupled to a bath,\cite{Rubtsov08} a problem for which spin is not conserved. While the exact summation of all contributions to all orders\cite{Iskakov16} would restore this symmetry, the perturbative ladder Dual Fermion vertex only partially cancels the $GG$ bubble contribution.
We expect $\omega_s(0,0)$ to decrease in energy as the system becomes metallic. This is reflected in both our high temperature and doped data. 
\section{Results}\label{sec:results}
We present DF results for the dynamical spin susceptibility of the single band Hubbard model obtained in the intermediate interaction strength regime where both weak coupling perturbative methods and strong coupling expansions fail.  We have shown that the DF results are qualitatively correct at high temperatures and note that our calculations have the advantage of a fine-grained momentum resolution of up to $64\times 64 = 4096$ points in the Brillouin zone.

\subsection{Spin-wave dispersion at half filling}
\begin{figure}
\includegraphics[width=\linewidth]{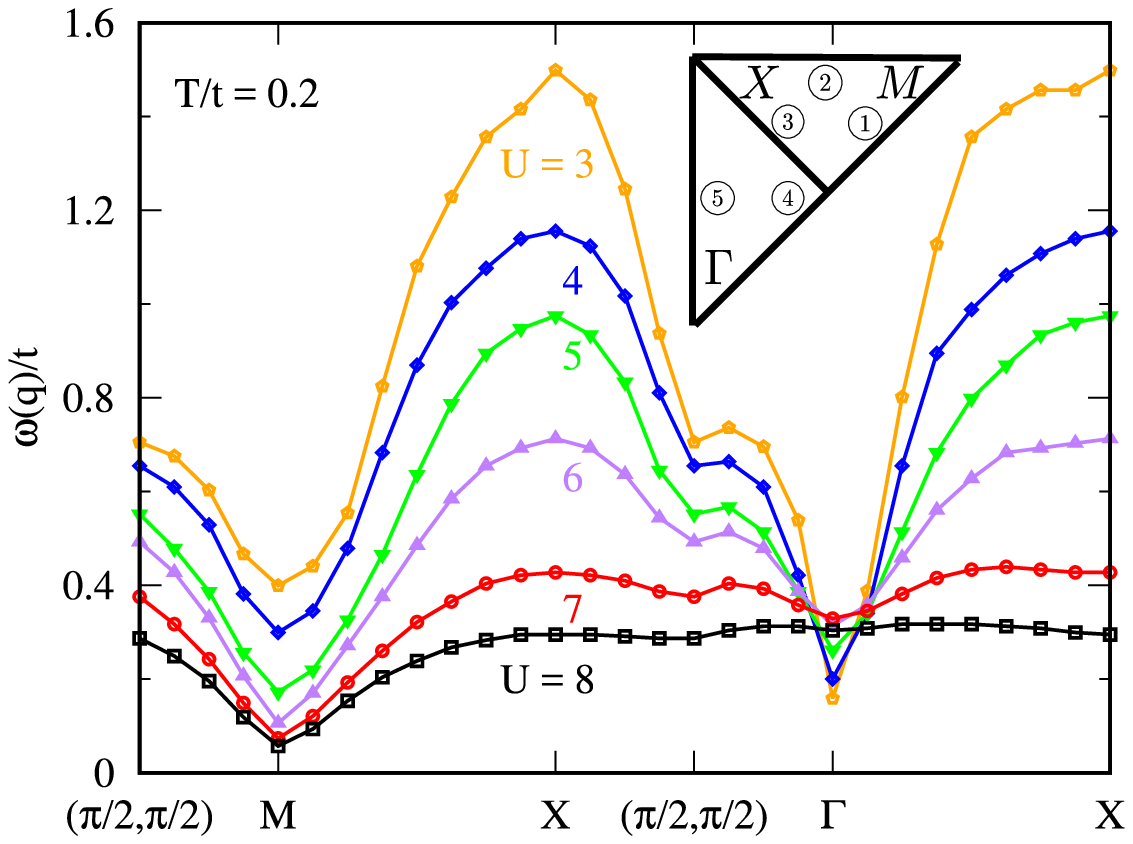}\\
\caption{\label{fig:vsU}  DF results for $\omega_s(q)/t$ with increasing $U/t=3\to 8$, at $T/t=0.2$, $t^\prime/t=-0.3,$ and $n=1$. Inset: path in the Brillouin zone.}
\end{figure}

Fig.~\ref{fig:vsU} shows the momentum dependence of the spin wave dispersion $\omega_s(q)$ from the DF method of the half-filled system along high symmetry paths in the first Brillouin zone (see inset), for interaction strengths ranging from $U/t=3$ to $U/t=8$. Results are obtained at temperature $T/t=0.2$ and next-nearest neighbor hopping $t'/t=-0.3$.  At weak interaction $U/t=3$ (orange circles), the spin mode, $\omega_s(q)$, at $M$ and $X$ points is at higher energy than other interaction strengths. Increasing the interaction towards $U/t=8$ results in a decrease in $\omega_s(q)$ at these two points.  This reduction in energy is accompanied by an increase in amplitude of the susceptibility, indicating strong low-energy spin fluctuations.

\begin{figure}
\includegraphics[width=\linewidth]{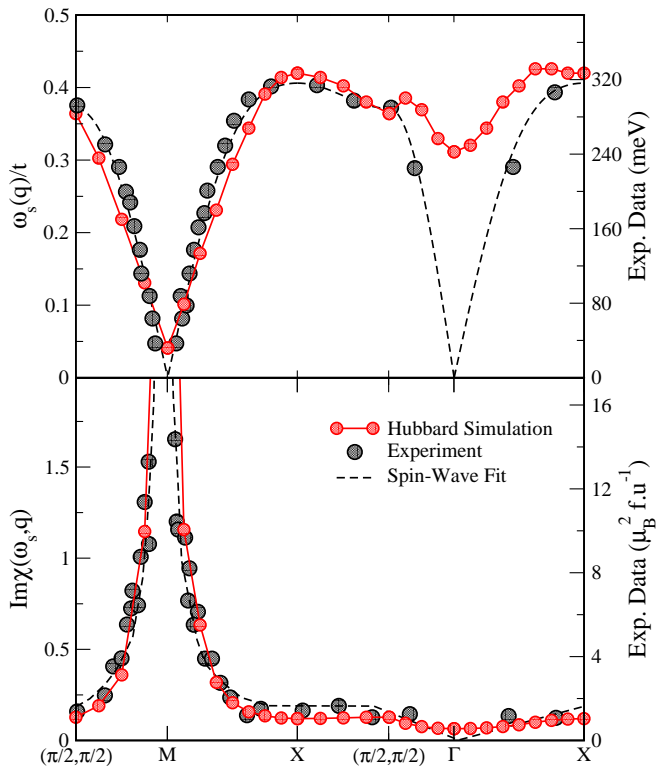}
\caption{\label{fig:exp}  Top: DF simulation results of  $\omega_s(q)/t$ for $U/t=7.6$ at $T/t=0.2$ with $t^{\prime}/t=-0.3$ at $\langle n \rangle =1$. Dashed line and filled grey circles are linear spin-wave fit and experimental data from Coldea et al. \cite{Coldea01}.  Left (theory) and right (experiment) axes differ by a prefactor of 2.36. Bottom: DF results for $\text{Im}\chi(\omega_s(q))$ and experimental curves rescaled to the Hubbard model.  Right hand axis is scale of original neutron data from \textcite{Coldea01}.
}
\end{figure}

We compare our numerical data to the experimental spin susceptibilities obtained on LaCuO$_4$\cite{Coldea01} (black points, right axis, see also Ref.~\onlinecite{Headings10}) in the upper panel of Fig.~\ref{fig:exp}. Also shown is a linear spin-wave fit with parameters determined by \textcite{Coldea01}
Our simulations were obtained for $U/t=7.6$ and $t'/t=-0.3$.
Ref.~\onlinecite{Coldea01} fitted the linear spin-wave results by a set of nearest and further Heisenberg exchange constants, $J,J',J''$ as well as a substantial ring exchange $J_c$, and found $J=138$meV, $J'=J''=2$meV and $J_c=40$meV.  

From a comparison between the overall shape of the spin excitation dispersion in experiment, Fig.~\ref{fig:exp}, and calculation, and in particular from the behavior between $M$, $X$, and $(\pi/2,\pi/2)$, we can conclude that an appropriate Hubbard interaction strength for modeling the data should be between $U=7t$ and $U=8t$. Specifically, the weak momentum dependence observed along the $M$-$X$-$(\pi/2,\pi/2)$ lines is inconsistent with an interaction strength much less than $U=7t$.

The lower panel of Fig.~\ref{fig:exp} shows calculated and experimental data for the imaginary part of the susceptibility at its maximum point $\omega_s(q)$. Spin excitations in both calculation and experiment are dominated by the strong low-energy excitation near $(\pi,\pi)$ and proportional to each other.

An obvious point of disagreement between our data for $\omega_s(q)$ and the experimental results is the overall magitude of the spin-wave dispersion. For generally accepted values of $U/t \sim 7-8$, and using a value of $t\sim0.35eV$,\cite{Andersen95} we find that $\omega_s(q=(\pi/2,\pi/2))=0.4t=142meV$. This value is a factor of $2.36$ below the experimental value.  Since we expect the DF method to be correct only up to a prefactor (see Fig.~\ref{fig:figDQMC_n} where $1/r\approx 2$ at the higher $T/t=1/3$) this scaling factor is consistent.

\begin{figure}
\includegraphics[width=0.8\linewidth]{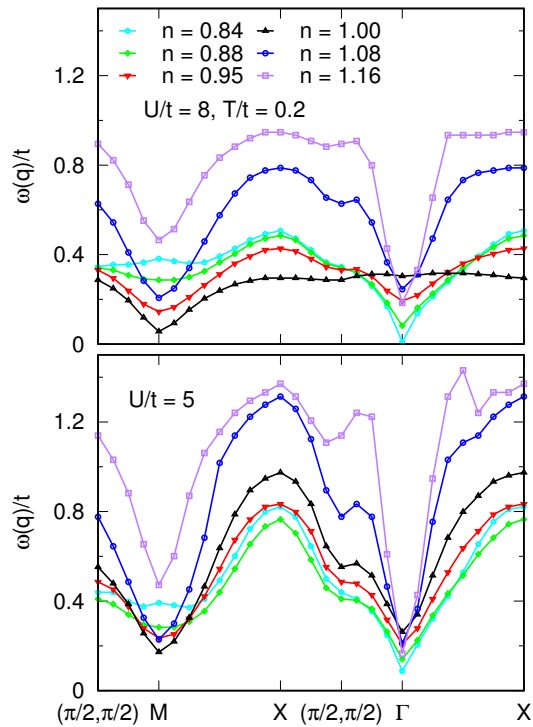}
\caption{\label{fig:Uvsn} Momentum dependence of spin-wave dispersion $\omega_s(q)/t$ for  densities indicated, for $U/t=8$ (upper panel) and $U/t=5$ (lower panel) at $T/t=0.2$ and $t^\prime/t=-0.3$.}
\end{figure}

\subsection{Doping dependence of the spin-wave dispersion}
The spin-wave dispersion is highly doping dependent (Fig.~\ref{fig:Uvsn}), and data shows a pronounced difference between electron and hole doping, with the spin wave energy increasing much faster for electron than for hole doping, except very near the $M$ point. The interpretation of DF data is complicated by the fact that the rescaling prefactor is strongly doping and interaction dependent (Fig.~\ref{fig:figDQMC_n}). Also, the temperature dependence of the rescaling factors is unknown (data is only available down to $T/t=1/3$). We show our results without rescaling $r$, but we expect the general conclusions we draw here to be robust based on $r$ at higher temperature. At both intermediate ($U/t=5$) and strong ($U/t=8$) interaction strength, $\omega_s(q)$ near $(\pi,\pi)$ quickly increases with doping. At the $X$ point, hole doping reduces $\omega_s$ and electron doping increases $\omega_s$ at intermediate interaction, whereas for strong coupling electron doping leads to a sharp increase and only a moderate increase and eventual saturation for hole doping. No change is visible upon hole doping at $(\pi/2,\pi/2)$, whereas electron doping increases $\omega_s(q)$ by a factor of at least two.

\begin{figure}
\includegraphics[width=\linewidth]{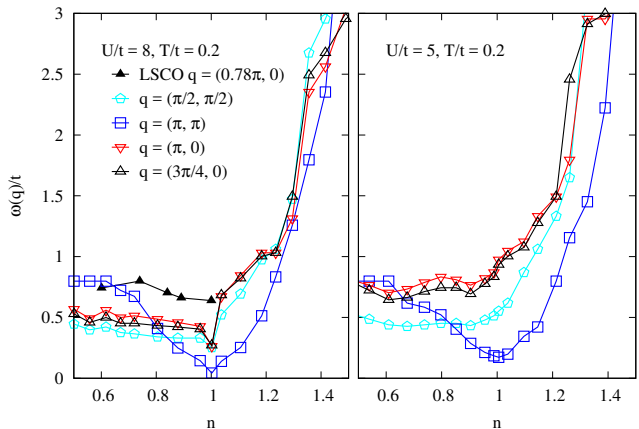}
\caption{\label{fig:qvsn} Spin-wave dispersion $\omega_s(q)$ for interaction strengths $U/t=8$ and $U/t=5$ at fixed $q$ values and $T/t=0.2$. 
 Also shown is experimental RIXS data on LSCO from \textcite{Dean13}.}
\end{figure}

\begin{figure}
\includegraphics[width=\linewidth]{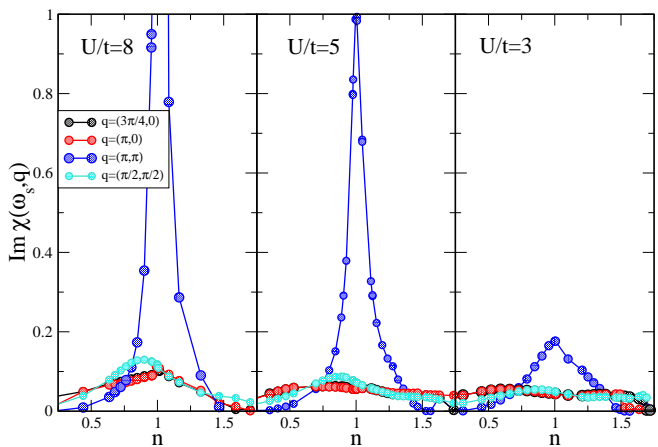}
\caption{\label{fig:magvsn} Amplitude of $\rm{Im}\chi(\omega_s,q)$ for fixed $q$ as a function of density for $U/t=8$, 5 and 3, at $t'/t=-0.3$ and $T/t=0.2$.}
\end{figure}

Fig.~\ref{fig:qvsn} shows a  detailed dependence of the spin-wave dispersion on doping, plotted as a function of occupation, $n$, for $U/t=8$ (left panel) and $U/t=5$ (right panel). In particular a strong difference between electron ($n>1$) and hole ($n<1$) doping is visible in the data at $(3\pi/4,0)$ and $(\pi,0)$, which shows no doping dependence on the hole-doped side but a rather strong momentum dependence on the electron-doped side. The data shown here are not rescaled to DQMC data. This result is qualitatively consistent with determinant quantum Monte Carlo calculations at higher temperature,\cite{Jia14} and with LSCO data from \textcite{Dean13} (see also Ref.~\onlinecite{Jia14}).
At $q=(\pi,\pi)$ both electron and hole doping increase $\omega_s$ shown in blue squares, which rises to higher energy for doping in each direction. 

Finally, as interaction strength decreases from $U/t=8$ to 5 we see that $q=(\pi,\pi)$ maintains a minimal value near $n=1$ but the mode energy increases; and that $q=(\pi,0)$ and $(3\pi/4,0)$ continues to show little change in hole doping and increases upon electron doping. 

Fig. \ref{fig:magvsn} shows the amplitude of $\rm{Im}\chi(\omega,q)$ for fixed $q$ as a function of density for $U/t = 8$, 5, and 3. The amplitudes of $\rm{Im}\chi(\omega,q)$ at $(3\pi/4,0)$, $(\pi,0)$, and $(\pi,\pi)$ are suppressed by both hole and electron dopings. For $q=(\pi/2,\pi/2)$, the amplitude of $\rm{Im}\chi(\omega,q)$ is enhanced by initial hole doping and then suppressed by doping more holes; on the electron doping side, the amplitude of $\rm{Im}\chi(\omega,q)$ is suppressed monotonically by electron doping at $q=(\pi/2,\pi/2)$.

\subsection{Spin excitation dispersion near M point}
\begin{figure}[tbh]
\includegraphics[width=0.95\linewidth]{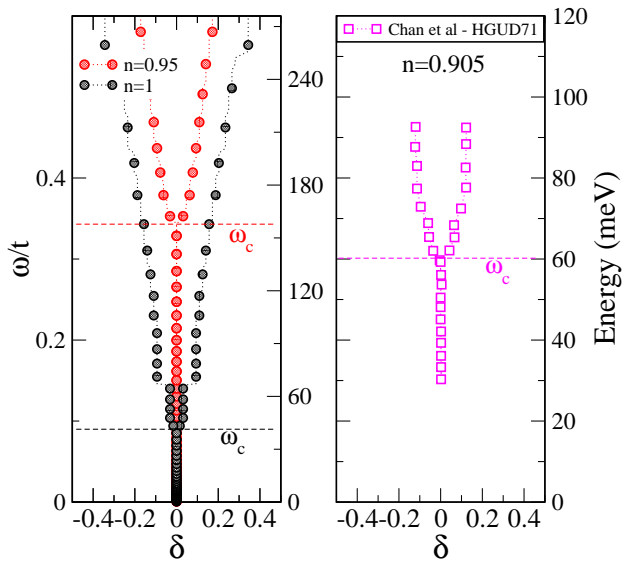}
\caption{\label{fig:greven} Left: Peaks in the spin excitation dispersion cuts of Im$\chi$ at fixed frequency for $U/t=8$, $t'/t=-0.3$, $T/t=0.2$ for densities $n=1,0.94$ and 0.86.  Right: Data from \textcite{Chan16}, their Fig.~4(a), showing data for HGUD71 ($n=0.905$).  The x-axis $\delta$ in both frames is the deviation from $(\pi,\pi)$ in reciprocal lattice units.    }
\end{figure}

The left panel of Fig.~\ref{fig:greven} shows a simulated spin excitation dispersion near $(\pi,\pi)$ for momenta $q=(\pi-\delta,\pi)$ and interaction strengths $U/t=8$ at half filling (black) and $n=0.95$ (red).  At low frequencies, the dispersion peaks are located at $\delta=0$.  At higher frequencies, the dispersions become incommensurate and begin to disperse at a characteristic frequency $\omega_c$, creating a `Y'-shape.  As the system is hole doped from $n=1$ to $n=0.95$, $\omega_c$ increases substantially.

Data are consistent with a resolution effect, where a single peak at the $(\pi,\pi)$ point (for $\omega=0$) with a constant width disperses linearly away from $(\pi,\pi)$ for higher frequencies.  Uncertainties of the analytical continuation procedure and the dual fermion approximation make it difficult to rule out alternative explanations.

For comparison, the right panel of Fig.~\ref{fig:greven} shows experimental data reproduced from Ref. \onlinecite{Chan16}.  A $t=460$~meV\cite{Das12,vishik:2014} allows us to establish the right hand axis of the left hand frame in Fig.~\ref{fig:greven}.  We can see that $\omega_c=40$~meV for $n=1$ is smaller than the experimental observation of $\omega_c=60$~meV while $n=0.95$ is much larger with $\omega_c=150$~meV.


\begin{figure}[tbh]
\includegraphics[width=0.7\linewidth]{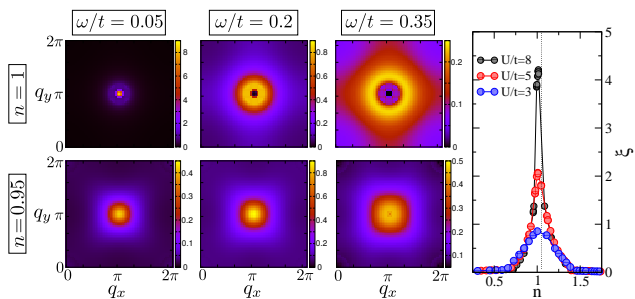}
\includegraphics[width=0.25\linewidth]{xi_vs_n_uvary_v2}
\caption{\label{fig:cplot} Left: Simulation results for $\rm{Im}\chi(\omega,q)$ at $U/t=8$, $t'/t=-0.3$ and $T/t=0.2$ at $n=1$ (top row) and $n=0.95$ (bottom row) for fixed frequencies in the $q_x$ and $q_y$ plane. Right: Doping dependence of the antiferromagnetic correlation length, $\xi$, for several interaction strengths $U$, obtained from a fit of $\chi$ along the direction $(q_x,q_y)=(q_x,\pi)$ with the function $f(q_x,\xi)=A/((q_x-\pi)^2+\xi^{-2})$.\cite{rohringer:2016,gukelberger:2016} \label{fig:xivsn} }
\end{figure}

To further examine the spin excitation spectrum around the $(\pi,\pi)$ region, we extract the correlation length $\xi$ by fitting the peak structure along the $(\pi,\delta)$ direction, see Fig.~\ref{fig:xivsn}, at $T/t=0.2$ as a function of doping for $U/t=8$, $5$, and $3$.  We observe that a reduction in interaction strength or doping in either direction causes a sharp decrease of the correlation length, which in turn results in a broadening of the spin structures at low energies. $\xi$ is marginally larger on the electron doped side (a result of the broken particle-hole symmetry) and this difference increases with the interaction strength $U/t$.

\section{Conclusions}\label{sec:discussion}
In summary, we have presented results from the Dual Fermion approximation for the spin excitation spectrum in the Hubbard model. The results show striking qualitative similarities with experiments in cuprate materials. The Dual Fermion results have been crosschecked both against DCA and against DQMC results at high temperature. Away from the $\Gamma$ point, the spectrum is believed to be accurate up to a momentum-independent prefactor and has a precise momentum resolution. As the prefactor is doping, interaction, and presumably temperature dependent, and as control calculations from DCA and DQMC are restricted to small systems or high temperature, a precise quantitative comparison between numerics and experimental spin excitations is not possible at this time. In addition, a quantitative comparison near the $\Gamma$ point requires a solution that respects spin conservation.


Nevertheless, from the overall shape of the spin excitation dispersion and in particular the behavior between $M$, $X$, and $(\pi/2,\pi/2)$, we can conclude that an appropriate Hubbard interaction strength for modeling experimental data should be between $U/t=7$ and $U/t=8$.

The existing literature shows that, on the single-particle level, agreement between model systems calculations and experiment is remarkable and is often not just qualitative but quantitative. On the two-particle level, similar progress has been made primarily for quantities that are either local (such as the NMR probe) or contain a slowly varying matrix element. One important exception is the work by Jia and collaborators \cite{Jia14} on simulations of the resonant $X$-ray scattering cross section. These authors presented $\chi''(q,\omega)$ obtained from DQMC calculations for the Hubbard model with $U=8t$ and $t'/t=-0.3$ at the temperature $T=t/3 \sim 1200K$, and reported quantitative agreement with magnetic neutron scattering measurements. Our DQMC results at these parameters are in quantitative agreement, as are our DF results (up to a rescaling); our studies of other $U$ values reinforce the conclusion that the Hubbard model with $U\sim 7-8t$ is a good description of cuprate physics. The continuous momentum resolution of the DF technique now shows that, in addition, many of the momentum dependent features seen in experiment are also observed in calculations.

\section{Acknowledgments}
We thank T. Devereaux and A.-M. Tremblay for fruitful discussions.
Dual Fermion methods development has been supported by the Simons collaboration on the many-electron problem, 
application to susceptibilities and neutron spectroscopy by NSF-DMR-1606348. 
JPFL acknowledges the support of the Natural Sciences and Engineering Research Council of Canada (NSERC), RGPIN-2017-04253. 
Computational resources were provided by TG-DMR130036, ACENET and Compute Canada.
\bibliographystyle{apsrev4-1}
\bibliography{refs.bib}

\begin{thebibliography}{67}%
\makeatletter
\providecommand \@ifxundefined [1]{%
 \@ifx{#1\undefined}
}%
\providecommand \@ifnum [1]{%
 \ifnum #1\expandafter \@firstoftwo
 \else \expandafter \@secondoftwo
 \fi
}%
\providecommand \@ifx [1]{%
 \ifx #1\expandafter \@firstoftwo
 \else \expandafter \@secondoftwo
 \fi
}%
\providecommand \natexlab [1]{#1}%
\providecommand \enquote  [1]{``#1''}%
\providecommand \bibnamefont  [1]{#1}%
\providecommand \bibfnamefont [1]{#1}%
\providecommand \citenamefont [1]{#1}%
\providecommand \href@noop [0]{\@secondoftwo}%
\providecommand \href [0]{\begingroup \@sanitize@url \@href}%
\providecommand \@href[1]{\@@startlink{#1}\@@href}%
\providecommand \@@href[1]{\endgroup#1\@@endlink}%
\providecommand \@sanitize@url [0]{\catcode `\\12\catcode `\$12\catcode
  `\&12\catcode `\#12\catcode `\^12\catcode `\_12\catcode `\%12\relax}%
\providecommand \@@startlink[1]{}%
\providecommand \@@endlink[0]{}%
\providecommand \url  [0]{\begingroup\@sanitize@url \@url }%
\providecommand \@url [1]{\endgroup\@href {#1}{\urlprefix }}%
\providecommand \urlprefix  [0]{URL }%
\providecommand \Eprint [0]{\href }%
\providecommand \doibase [0]{http://dx.doi.org/}%
\providecommand \selectlanguage [0]{\@gobble}%
\providecommand \bibinfo  [0]{\@secondoftwo}%
\providecommand \bibfield  [0]{\@secondoftwo}%
\providecommand \translation [1]{[#1]}%
\providecommand \BibitemOpen [0]{}%
\providecommand \bibitemStop [0]{}%
\providecommand \bibitemNoStop [0]{.\EOS\space}%
\providecommand \EOS [0]{\spacefactor3000\relax}%
\providecommand \BibitemShut  [1]{\csname bibitem#1\endcsname}%
\let\auto@bib@innerbib\@empty
\bibitem [{\citenamefont {Fujita}\ \emph {et~al.}(2012)\citenamefont {Fujita},
  \citenamefont {Hiraka}, \citenamefont {Matsuda}, \citenamefont {Matsuura},
  \citenamefont {Tranquada}, \citenamefont {Wakimoto}, \citenamefont {Xu},\
  and\ \citenamefont {Yamada}}]{yamada:2012}%
  \BibitemOpen
  \bibfield  {author} {\bibinfo {author} {\bibfnamefont {M.}~\bibnamefont
  {Fujita}}, \bibinfo {author} {\bibfnamefont {H.}~\bibnamefont {Hiraka}},
  \bibinfo {author} {\bibfnamefont {M.}~\bibnamefont {Matsuda}}, \bibinfo
  {author} {\bibfnamefont {M.}~\bibnamefont {Matsuura}}, \bibinfo {author}
  {\bibfnamefont {J.~M.}\ \bibnamefont {Tranquada}}, \bibinfo {author}
  {\bibfnamefont {S.}~\bibnamefont {Wakimoto}}, \bibinfo {author}
  {\bibfnamefont {G.}~\bibnamefont {Xu}}, \ and\ \bibinfo {author}
  {\bibfnamefont {K.}~\bibnamefont {Yamada}},\ }\href {\doibase
  10.1143/JPSJ.81.011007} {\bibfield  {journal} {\bibinfo  {journal} {Journal
  of the Physical Society of Japan}\ }\textbf {\bibinfo {volume} {81}},\
  \bibinfo {pages} {011007} (\bibinfo {year} {2012})},\ \Eprint
  {http://arxiv.org/abs/http://dx.doi.org/10.1143/JPSJ.81.011007}
  {http://dx.doi.org/10.1143/JPSJ.81.011007} \BibitemShut {NoStop}%
\bibitem [{\citenamefont {Coldea}\ \emph {et~al.}(2001)\citenamefont {Coldea},
  \citenamefont {Hayden}, \citenamefont {Aeppli}, \citenamefont {Perring},
  \citenamefont {Frost}, \citenamefont {Mason}, \citenamefont {Cheong},\ and\
  \citenamefont {Fisk}}]{Coldea01}%
  \BibitemOpen
  \bibfield  {author} {\bibinfo {author} {\bibfnamefont {R.}~\bibnamefont
  {Coldea}}, \bibinfo {author} {\bibfnamefont {S.~M.}\ \bibnamefont {Hayden}},
  \bibinfo {author} {\bibfnamefont {G.}~\bibnamefont {Aeppli}}, \bibinfo
  {author} {\bibfnamefont {T.~G.}\ \bibnamefont {Perring}}, \bibinfo {author}
  {\bibfnamefont {C.~D.}\ \bibnamefont {Frost}}, \bibinfo {author}
  {\bibfnamefont {T.~E.}\ \bibnamefont {Mason}}, \bibinfo {author}
  {\bibfnamefont {S.-W.}\ \bibnamefont {Cheong}}, \ and\ \bibinfo {author}
  {\bibfnamefont {Z.}~\bibnamefont {Fisk}},\ }\href {\doibase
  10.1103/PhysRevLett.86.5377} {\bibfield  {journal} {\bibinfo  {journal}
  {Phys. Rev. Lett.}\ }\textbf {\bibinfo {volume} {86}},\ \bibinfo {pages}
  {5377} (\bibinfo {year} {2001})}\BibitemShut {NoStop}%
\bibitem [{\citenamefont {Lipscombe}\ \emph {et~al.}(2009)\citenamefont
  {Lipscombe}, \citenamefont {Vignolle}, \citenamefont {Perring}, \citenamefont
  {Frost},\ and\ \citenamefont {Hayden}}]{Lipscombe09}%
  \BibitemOpen
  \bibfield  {author} {\bibinfo {author} {\bibfnamefont {O.~J.}\ \bibnamefont
  {Lipscombe}}, \bibinfo {author} {\bibfnamefont {B.}~\bibnamefont {Vignolle}},
  \bibinfo {author} {\bibfnamefont {T.~G.}\ \bibnamefont {Perring}}, \bibinfo
  {author} {\bibfnamefont {C.~D.}\ \bibnamefont {Frost}}, \ and\ \bibinfo
  {author} {\bibfnamefont {S.~M.}\ \bibnamefont {Hayden}},\ }\href {\doibase
  10.1103/PhysRevLett.102.167002} {\bibfield  {journal} {\bibinfo  {journal}
  {Phys. Rev. Lett.}\ }\textbf {\bibinfo {volume} {102}},\ \bibinfo {pages}
  {167002} (\bibinfo {year} {2009})}\BibitemShut {NoStop}%
\bibitem [{\citenamefont {Lipscombe}\ \emph {et~al.}(2007)\citenamefont
  {Lipscombe}, \citenamefont {Hayden}, \citenamefont {Vignolle}, \citenamefont
  {McMorrow},\ and\ \citenamefont {Perring}}]{lipscombe:2007}%
  \BibitemOpen
  \bibfield  {author} {\bibinfo {author} {\bibfnamefont {O.~J.}\ \bibnamefont
  {Lipscombe}}, \bibinfo {author} {\bibfnamefont {S.~M.}\ \bibnamefont
  {Hayden}}, \bibinfo {author} {\bibfnamefont {B.}~\bibnamefont {Vignolle}},
  \bibinfo {author} {\bibfnamefont {D.~F.}\ \bibnamefont {McMorrow}}, \ and\
  \bibinfo {author} {\bibfnamefont {T.~G.}\ \bibnamefont {Perring}},\ }\href
  {\doibase 10.1103/PhysRevLett.99.067002} {\bibfield  {journal} {\bibinfo
  {journal} {Phys. Rev. Lett.}\ }\textbf {\bibinfo {volume} {99}},\ \bibinfo
  {pages} {067002} (\bibinfo {year} {2007})}\BibitemShut {NoStop}%
\bibitem [{\citenamefont {Vignolle}\ \emph {et~al.}(2007)\citenamefont
  {Vignolle}, \citenamefont {Hayden}, \citenamefont {McMorrow}, \citenamefont
  {Ronnow}, \citenamefont {Lake}, \citenamefont {Frost},\ and\ \citenamefont
  {Perring}}]{Vignolle07}%
  \BibitemOpen
  \bibfield  {author} {\bibinfo {author} {\bibfnamefont {B.}~\bibnamefont
  {Vignolle}}, \bibinfo {author} {\bibfnamefont {S.~M.}\ \bibnamefont
  {Hayden}}, \bibinfo {author} {\bibfnamefont {D.~F.}\ \bibnamefont
  {McMorrow}}, \bibinfo {author} {\bibfnamefont {H.~M.}\ \bibnamefont
  {Ronnow}}, \bibinfo {author} {\bibfnamefont {B.}~\bibnamefont {Lake}},
  \bibinfo {author} {\bibfnamefont {C.~D.}\ \bibnamefont {Frost}}, \ and\
  \bibinfo {author} {\bibfnamefont {T.~G.}\ \bibnamefont {Perring}},\ }\href
  {\doibase 10.1038/nphys546} {\bibfield  {journal} {\bibinfo  {journal} {Nat
  Phys}\ }\textbf {\bibinfo {volume} {3}},\ \bibinfo {pages} {163} (\bibinfo
  {year} {2007})}\BibitemShut {NoStop}%
\bibitem [{\citenamefont {Shekhter}\ \emph {et~al.}(2013)\citenamefont
  {Shekhter}, \citenamefont {Ramshaw}, \citenamefont {Liang}, \citenamefont
  {Hardy}, \citenamefont {Bonn}, \citenamefont {Balakirev}, \citenamefont
  {McDonald}, \citenamefont {Betts}, \citenamefont {Riggs},\ and\ \citenamefont
  {Migliori}}]{shekhter:2013}%
  \BibitemOpen
  \bibfield  {author} {\bibinfo {author} {\bibfnamefont {A.}~\bibnamefont
  {Shekhter}}, \bibinfo {author} {\bibfnamefont {B.~J.}\ \bibnamefont
  {Ramshaw}}, \bibinfo {author} {\bibfnamefont {R.}~\bibnamefont {Liang}},
  \bibinfo {author} {\bibfnamefont {W.~N.}\ \bibnamefont {Hardy}}, \bibinfo
  {author} {\bibfnamefont {D.~A.}\ \bibnamefont {Bonn}}, \bibinfo {author}
  {\bibfnamefont {F.~F.}\ \bibnamefont {Balakirev}}, \bibinfo {author}
  {\bibfnamefont {R.~D.}\ \bibnamefont {McDonald}}, \bibinfo {author}
  {\bibfnamefont {J.~B.}\ \bibnamefont {Betts}}, \bibinfo {author}
  {\bibfnamefont {S.~C.}\ \bibnamefont {Riggs}}, \ and\ \bibinfo {author}
  {\bibfnamefont {A.}~\bibnamefont {Migliori}},\ }\href@noop {} {\bibfield
  {journal} {\bibinfo  {journal} {Nature}\ }\textbf {\bibinfo {volume} {498}},\
  \bibinfo {pages} {75} (\bibinfo {year} {2013})}\BibitemShut {NoStop}%
\bibitem [{\citenamefont {LeBoeuf}\ \emph {et~al.}(2013)\citenamefont
  {LeBoeuf}, \citenamefont {Kramer}, \citenamefont {Hardy}, \citenamefont
  {Liang}, \citenamefont {Bonn},\ and\ \citenamefont {Proust}}]{leboeuf:2013}%
  \BibitemOpen
  \bibfield  {author} {\bibinfo {author} {\bibfnamefont {D.}~\bibnamefont
  {LeBoeuf}}, \bibinfo {author} {\bibfnamefont {S.}~\bibnamefont {Kramer}},
  \bibinfo {author} {\bibfnamefont {W.~N.}\ \bibnamefont {Hardy}}, \bibinfo
  {author} {\bibfnamefont {R.}~\bibnamefont {Liang}}, \bibinfo {author}
  {\bibfnamefont {D.~A.}\ \bibnamefont {Bonn}}, \ and\ \bibinfo {author}
  {\bibfnamefont {C.}~\bibnamefont {Proust}},\ }\href {\doibase
  10.1038/nphys2502} {\bibfield  {journal} {\bibinfo  {journal} {Nat Phys}\
  }\textbf {\bibinfo {volume} {9}},\ \bibinfo {pages} {79} (\bibinfo {year}
  {2013})}\BibitemShut {NoStop}%
\bibitem [{\citenamefont {Grissonnanche}\ \emph {et~al.}(2015)\citenamefont
  {Grissonnanche}, \citenamefont {Laliberte}, \citenamefont
  {Dufour-Beausejour}, \citenamefont {Riopel}, \citenamefont {Badoux},
  \citenamefont {Caouette-Mansour}, \citenamefont {Matusiak}, \citenamefont
  {Juneau-Fecteau}, \citenamefont {Bourgeois-Hope}, \citenamefont
  {Cyr-Choiniere}, \citenamefont {Baglo}, \citenamefont {Ramshaw},
  \citenamefont {Liang}, \citenamefont {Bonn}, \citenamefont {Hardy},
  \citenamefont {Kramer}, \citenamefont {LeBoeuf}, \citenamefont {Graf},
  \citenamefont {Doiron-Leyraud},\ and\ \citenamefont
  {Taillefer}}]{grissonnanche:2015}%
  \BibitemOpen
  \bibfield  {author} {\bibinfo {author} {\bibfnamefont {G.}~\bibnamefont
  {Grissonnanche}}, \bibinfo {author} {\bibfnamefont {F.}~\bibnamefont
  {Laliberte}}, \bibinfo {author} {\bibfnamefont {S.}~\bibnamefont
  {Dufour-Beausejour}}, \bibinfo {author} {\bibfnamefont {A.}~\bibnamefont
  {Riopel}}, \bibinfo {author} {\bibfnamefont {S.}~\bibnamefont {Badoux}},
  \bibinfo {author} {\bibfnamefont {M.}~\bibnamefont {Caouette-Mansour}},
  \bibinfo {author} {\bibfnamefont {M.}~\bibnamefont {Matusiak}}, \bibinfo
  {author} {\bibfnamefont {A.}~\bibnamefont {Juneau-Fecteau}}, \bibinfo
  {author} {\bibfnamefont {P.}~\bibnamefont {Bourgeois-Hope}}, \bibinfo
  {author} {\bibfnamefont {O.}~\bibnamefont {Cyr-Choiniere}}, \bibinfo {author}
  {\bibfnamefont {J.~C.}\ \bibnamefont {Baglo}}, \bibinfo {author}
  {\bibfnamefont {B.~J.}\ \bibnamefont {Ramshaw}}, \bibinfo {author}
  {\bibfnamefont {R.}~\bibnamefont {Liang}}, \bibinfo {author} {\bibfnamefont
  {D.~A.}\ \bibnamefont {Bonn}}, \bibinfo {author} {\bibfnamefont {W.~N.}\
  \bibnamefont {Hardy}}, \bibinfo {author} {\bibfnamefont {S.}~\bibnamefont
  {Kramer}}, \bibinfo {author} {\bibfnamefont {D.}~\bibnamefont {LeBoeuf}},
  \bibinfo {author} {\bibfnamefont {D.}~\bibnamefont {Graf}}, \bibinfo {author}
  {\bibfnamefont {N.}~\bibnamefont {Doiron-Leyraud}}, \ and\ \bibinfo {author}
  {\bibfnamefont {L.}~\bibnamefont {Taillefer}},\ }\href@noop {} {\bibfield
  {journal} {\bibinfo  {journal} {arXiv}\ ,\ \bibinfo {pages} {1508.05486}}
  (\bibinfo {year} {2015})}\BibitemShut {NoStop}%
\bibitem [{\citenamefont {Badoux}\ \emph {et~al.}(2016)\citenamefont {Badoux},
  \citenamefont {Afshar}, \citenamefont {Michon}, \citenamefont {Ouellet},
  \citenamefont {Fortier}, \citenamefont {LeBoeuf}, \citenamefont {Croft},
  \citenamefont {Lester}, \citenamefont {Hayden}, \citenamefont {Takagi},
  \citenamefont {Yamada}, \citenamefont {Graf}, \citenamefont
  {Doiron-Leyraud},\ and\ \citenamefont {Taillefer}}]{badoux:2016}%
  \BibitemOpen
  \bibfield  {author} {\bibinfo {author} {\bibfnamefont {S.}~\bibnamefont
  {Badoux}}, \bibinfo {author} {\bibfnamefont {S.~A.~A.}\ \bibnamefont
  {Afshar}}, \bibinfo {author} {\bibfnamefont {B.}~\bibnamefont {Michon}},
  \bibinfo {author} {\bibfnamefont {A.}~\bibnamefont {Ouellet}}, \bibinfo
  {author} {\bibfnamefont {S.}~\bibnamefont {Fortier}}, \bibinfo {author}
  {\bibfnamefont {D.}~\bibnamefont {LeBoeuf}}, \bibinfo {author} {\bibfnamefont
  {T.~P.}\ \bibnamefont {Croft}}, \bibinfo {author} {\bibfnamefont
  {C.}~\bibnamefont {Lester}}, \bibinfo {author} {\bibfnamefont {S.~M.}\
  \bibnamefont {Hayden}}, \bibinfo {author} {\bibfnamefont {H.}~\bibnamefont
  {Takagi}}, \bibinfo {author} {\bibfnamefont {K.}~\bibnamefont {Yamada}},
  \bibinfo {author} {\bibfnamefont {D.}~\bibnamefont {Graf}}, \bibinfo {author}
  {\bibfnamefont {N.}~\bibnamefont {Doiron-Leyraud}}, \ and\ \bibinfo {author}
  {\bibfnamefont {L.}~\bibnamefont {Taillefer}},\ }\href {\doibase
  10.1103/PhysRevX.6.021004} {\bibfield  {journal} {\bibinfo  {journal} {Phys.
  Rev. X}\ }\textbf {\bibinfo {volume} {6}},\ \bibinfo {pages} {021004}
  (\bibinfo {year} {2016})}\BibitemShut {NoStop}%
\bibitem [{\citenamefont {{Le Tacon}}\ \emph {et~al.}(2011)\citenamefont {{Le
  Tacon}}, \citenamefont {Ghiringhelli}, \citenamefont {Chaloupka},
  \citenamefont {{Moretti Sala}}, \citenamefont {Hinkov}, \citenamefont
  {Haverkort}, \citenamefont {Minola}, \citenamefont {Bakr}, \citenamefont
  {Zhou}, \citenamefont {Blanco-Canosa}, \citenamefont {Monney}, \citenamefont
  {Song}, \citenamefont {Sun}, \citenamefont {Lin}, \citenamefont {{De Luca}},
  \citenamefont {Salluzzo}, \citenamefont {Khaliullin}, \citenamefont
  {Schmitt}, \citenamefont {Braicovich},\ and\ \citenamefont
  {Keimer}}]{letacon:rixs}%
  \BibitemOpen
  \bibfield  {author} {\bibinfo {author} {\bibfnamefont {M.}~\bibnamefont {{Le
  Tacon}}}, \bibinfo {author} {\bibfnamefont {G.}~\bibnamefont {Ghiringhelli}},
  \bibinfo {author} {\bibfnamefont {J.}~\bibnamefont {Chaloupka}}, \bibinfo
  {author} {\bibfnamefont {M.}~\bibnamefont {{Moretti Sala}}}, \bibinfo
  {author} {\bibfnamefont {V.}~\bibnamefont {Hinkov}}, \bibinfo {author}
  {\bibfnamefont {M.~W.}\ \bibnamefont {Haverkort}}, \bibinfo {author}
  {\bibfnamefont {M.}~\bibnamefont {Minola}}, \bibinfo {author} {\bibfnamefont
  {M.}~\bibnamefont {Bakr}}, \bibinfo {author} {\bibfnamefont {K.~J.}\
  \bibnamefont {Zhou}}, \bibinfo {author} {\bibfnamefont {S.}~\bibnamefont
  {Blanco-Canosa}}, \bibinfo {author} {\bibfnamefont {C.}~\bibnamefont
  {Monney}}, \bibinfo {author} {\bibfnamefont {Y.~T.}\ \bibnamefont {Song}},
  \bibinfo {author} {\bibfnamefont {G.~L.}\ \bibnamefont {Sun}}, \bibinfo
  {author} {\bibfnamefont {C.~T.}\ \bibnamefont {Lin}}, \bibinfo {author}
  {\bibfnamefont {G.~M.}\ \bibnamefont {{De Luca}}}, \bibinfo {author}
  {\bibfnamefont {M.}~\bibnamefont {Salluzzo}}, \bibinfo {author}
  {\bibfnamefont {G.}~\bibnamefont {Khaliullin}}, \bibinfo {author}
  {\bibfnamefont {T.}~\bibnamefont {Schmitt}}, \bibinfo {author} {\bibfnamefont
  {L.}~\bibnamefont {Braicovich}}, \ and\ \bibinfo {author} {\bibfnamefont
  {B.}~\bibnamefont {Keimer}},\ }\href@noop {} {\bibfield  {journal} {\bibinfo
  {journal} {Nature Physics}\ }\textbf {\bibinfo {volume} {7}},\ \bibinfo
  {pages} {725} (\bibinfo {year} {2011})}\BibitemShut {NoStop}%
\bibitem [{\citenamefont {Braicovich}\ \emph {et~al.}(2010)\citenamefont
  {Braicovich}, \citenamefont {van~den Brink}, \citenamefont {Bisogni},
  \citenamefont {Sala}, \citenamefont {Ament}, \citenamefont {Brookes},
  \citenamefont {De~Luca}, \citenamefont {Salluzzo}, \citenamefont {Schmitt},
  \citenamefont {Strocov},\ and\ \citenamefont {Ghiringhelli}}]{Braicovich10}%
  \BibitemOpen
  \bibfield  {author} {\bibinfo {author} {\bibfnamefont {L.}~\bibnamefont
  {Braicovich}}, \bibinfo {author} {\bibfnamefont {J.}~\bibnamefont {van~den
  Brink}}, \bibinfo {author} {\bibfnamefont {V.}~\bibnamefont {Bisogni}},
  \bibinfo {author} {\bibfnamefont {M.~M.}\ \bibnamefont {Sala}}, \bibinfo
  {author} {\bibfnamefont {L.~J.~P.}\ \bibnamefont {Ament}}, \bibinfo {author}
  {\bibfnamefont {N.~B.}\ \bibnamefont {Brookes}}, \bibinfo {author}
  {\bibfnamefont {G.~M.}\ \bibnamefont {De~Luca}}, \bibinfo {author}
  {\bibfnamefont {M.}~\bibnamefont {Salluzzo}}, \bibinfo {author}
  {\bibfnamefont {T.}~\bibnamefont {Schmitt}}, \bibinfo {author} {\bibfnamefont
  {V.~N.}\ \bibnamefont {Strocov}}, \ and\ \bibinfo {author} {\bibfnamefont
  {G.}~\bibnamefont {Ghiringhelli}},\ }\href {\doibase
  10.1103/PhysRevLett.104.077002} {\bibfield  {journal} {\bibinfo  {journal}
  {Phys. Rev. Lett.}\ }\textbf {\bibinfo {volume} {104}},\ \bibinfo {pages}
  {077002} (\bibinfo {year} {2010})}\BibitemShut {NoStop}%
\bibitem [{\citenamefont {Devereaux}\ and\ \citenamefont
  {Hackl}(2007)}]{Devereaux07}%
  \BibitemOpen
  \bibfield  {author} {\bibinfo {author} {\bibfnamefont {T.~P.}\ \bibnamefont
  {Devereaux}}\ and\ \bibinfo {author} {\bibfnamefont {R.}~\bibnamefont
  {Hackl}},\ }\href {\doibase 10.1103/RevModPhys.79.175} {\bibfield  {journal}
  {\bibinfo  {journal} {Rev. Mod. Phys.}\ }\textbf {\bibinfo {volume} {79}},\
  \bibinfo {pages} {175} (\bibinfo {year} {2007})}\BibitemShut {NoStop}%
\bibitem [{\citenamefont {Basov}\ and\ \citenamefont {Timusk}(2005)}]{Basov05}%
  \BibitemOpen
  \bibfield  {author} {\bibinfo {author} {\bibfnamefont {D.~N.}\ \bibnamefont
  {Basov}}\ and\ \bibinfo {author} {\bibfnamefont {T.}~\bibnamefont {Timusk}},\
  }\href {\doibase 10.1103/RevModPhys.77.721} {\bibfield  {journal} {\bibinfo
  {journal} {Rev. Mod. Phys.}\ }\textbf {\bibinfo {volume} {77}},\ \bibinfo
  {pages} {721} (\bibinfo {year} {2005})}\BibitemShut {NoStop}%
\bibitem [{\citenamefont {Huscroft}\ \emph {et~al.}(2001)\citenamefont
  {Huscroft}, \citenamefont {Jarrell}, \citenamefont {Maier}, \citenamefont
  {Moukouri},\ and\ \citenamefont {Tahvildarzadeh}}]{Huscroft01}%
  \BibitemOpen
  \bibfield  {author} {\bibinfo {author} {\bibfnamefont {C.}~\bibnamefont
  {Huscroft}}, \bibinfo {author} {\bibfnamefont {M.}~\bibnamefont {Jarrell}},
  \bibinfo {author} {\bibfnamefont {T.}~\bibnamefont {Maier}}, \bibinfo
  {author} {\bibfnamefont {S.}~\bibnamefont {Moukouri}}, \ and\ \bibinfo
  {author} {\bibfnamefont {A.~N.}\ \bibnamefont {Tahvildarzadeh}},\ }\href
  {\doibase 10.1103/PhysRevLett.86.139} {\bibfield  {journal} {\bibinfo
  {journal} {Phys. Rev. Lett.}\ }\textbf {\bibinfo {volume} {86}},\ \bibinfo
  {pages} {139} (\bibinfo {year} {2001})}\BibitemShut {NoStop}%
\bibitem [{\citenamefont {Parcollet}\ \emph {et~al.}(2004)\citenamefont
  {Parcollet}, \citenamefont {Biroli},\ and\ \citenamefont
  {Kotliar}}]{Parcollet04}%
  \BibitemOpen
  \bibfield  {author} {\bibinfo {author} {\bibfnamefont {O.}~\bibnamefont
  {Parcollet}}, \bibinfo {author} {\bibfnamefont {G.}~\bibnamefont {Biroli}}, \
  and\ \bibinfo {author} {\bibfnamefont {G.}~\bibnamefont {Kotliar}},\ }\href
  {\doibase 10.1103/PhysRevLett.92.226402} {\bibfield  {journal} {\bibinfo
  {journal} {Phys. Rev. Lett.}\ }\textbf {\bibinfo {volume} {92}},\ \bibinfo
  {pages} {226402} (\bibinfo {year} {2004})}\BibitemShut {NoStop}%
\bibitem [{\citenamefont {Macridin}\ \emph {et~al.}(2006)\citenamefont
  {Macridin}, \citenamefont {Jarrell}, \citenamefont {Maier}, \citenamefont
  {Kent},\ and\ \citenamefont {D'Azevedo}}]{Macridin06}%
  \BibitemOpen
  \bibfield  {author} {\bibinfo {author} {\bibfnamefont {A.}~\bibnamefont
  {Macridin}}, \bibinfo {author} {\bibfnamefont {M.}~\bibnamefont {Jarrell}},
  \bibinfo {author} {\bibfnamefont {T.}~\bibnamefont {Maier}}, \bibinfo
  {author} {\bibfnamefont {P.~R.~C.}\ \bibnamefont {Kent}}, \ and\ \bibinfo
  {author} {\bibfnamefont {E.}~\bibnamefont {D'Azevedo}},\ }\href {\doibase
  10.1103/PhysRevLett.97.036401} {\bibfield  {journal} {\bibinfo  {journal}
  {Phys. Rev. Lett.}\ }\textbf {\bibinfo {volume} {97}},\ \bibinfo {pages}
  {036401} (\bibinfo {year} {2006})}\BibitemShut {NoStop}%
\bibitem [{\citenamefont {Kyung}\ \emph {et~al.}(2006)\citenamefont {Kyung},
  \citenamefont {Kancharla}, \citenamefont {S\'en\'echal}, \citenamefont
  {Tremblay}, \citenamefont {Civelli},\ and\ \citenamefont
  {Kotliar}}]{Kyung06}%
  \BibitemOpen
  \bibfield  {author} {\bibinfo {author} {\bibfnamefont {B.}~\bibnamefont
  {Kyung}}, \bibinfo {author} {\bibfnamefont {S.~S.}\ \bibnamefont
  {Kancharla}}, \bibinfo {author} {\bibfnamefont {D.}~\bibnamefont
  {S\'en\'echal}}, \bibinfo {author} {\bibfnamefont {A.-M.~S.}\ \bibnamefont
  {Tremblay}}, \bibinfo {author} {\bibfnamefont {M.}~\bibnamefont {Civelli}}, \
  and\ \bibinfo {author} {\bibfnamefont {G.}~\bibnamefont {Kotliar}},\ }\href
  {\doibase 10.1103/PhysRevB.73.165114} {\bibfield  {journal} {\bibinfo
  {journal} {Phys. Rev. B}\ }\textbf {\bibinfo {volume} {73}},\ \bibinfo
  {pages} {165114} (\bibinfo {year} {2006})}\BibitemShut {NoStop}%
\bibitem [{\citenamefont {Werner}\ \emph {et~al.}(2009)\citenamefont {Werner},
  \citenamefont {Gull}, \citenamefont {Parcollet},\ and\ \citenamefont
  {Millis}}]{Werner098site}%
  \BibitemOpen
  \bibfield  {author} {\bibinfo {author} {\bibfnamefont {P.}~\bibnamefont
  {Werner}}, \bibinfo {author} {\bibfnamefont {E.}~\bibnamefont {Gull}},
  \bibinfo {author} {\bibfnamefont {O.}~\bibnamefont {Parcollet}}, \ and\
  \bibinfo {author} {\bibfnamefont {A.~J.}\ \bibnamefont {Millis}},\ }\href
  {\doibase 10.1103/PhysRevB.80.045120} {\bibfield  {journal} {\bibinfo
  {journal} {Phys. Rev. B}\ }\textbf {\bibinfo {volume} {80}},\ \bibinfo {eid}
  {045120} (\bibinfo {year} {2009})}\BibitemShut {NoStop}%
\bibitem [{\citenamefont {Gull}\ \emph {et~al.}(2009)\citenamefont {Gull},
  \citenamefont {Parcollet}, \citenamefont {Werner},\ and\ \citenamefont
  {Millis}}]{Gull09_8site}%
  \BibitemOpen
  \bibfield  {author} {\bibinfo {author} {\bibfnamefont {E.}~\bibnamefont
  {Gull}}, \bibinfo {author} {\bibfnamefont {O.}~\bibnamefont {Parcollet}},
  \bibinfo {author} {\bibfnamefont {P.}~\bibnamefont {Werner}}, \ and\ \bibinfo
  {author} {\bibfnamefont {A.~J.}\ \bibnamefont {Millis}},\ }\href {\doibase
  10.1103/PhysRevB.80.245102} {\bibfield  {journal} {\bibinfo  {journal} {Phys.
  Rev. B}\ }\textbf {\bibinfo {volume} {80}},\ \bibinfo {pages} {245102}
  (\bibinfo {year} {2009})}\BibitemShut {NoStop}%
\bibitem [{\citenamefont {Lin}\ \emph {et~al.}(2009)\citenamefont {Lin},
  \citenamefont {Gull},\ and\ \citenamefont {Millis}}]{Lin09}%
  \BibitemOpen
  \bibfield  {author} {\bibinfo {author} {\bibfnamefont {N.}~\bibnamefont
  {Lin}}, \bibinfo {author} {\bibfnamefont {E.}~\bibnamefont {Gull}}, \ and\
  \bibinfo {author} {\bibfnamefont {A.~J.}\ \bibnamefont {Millis}},\ }\href
  {\doibase 10.1103/PhysRevB.80.161105} {\bibfield  {journal} {\bibinfo
  {journal} {Phys. Rev. B}\ }\textbf {\bibinfo {volume} {80}},\ \bibinfo
  {pages} {161105} (\bibinfo {year} {2009})}\BibitemShut {NoStop}%
\bibitem [{\citenamefont {Liebsch}\ and\ \citenamefont
  {Tong}(2009)}]{Liebsch09}%
  \BibitemOpen
  \bibfield  {author} {\bibinfo {author} {\bibfnamefont {A.}~\bibnamefont
  {Liebsch}}\ and\ \bibinfo {author} {\bibfnamefont {N.-H.}\ \bibnamefont
  {Tong}},\ }\href {\doibase 10.1103/PhysRevB.80.165126} {\bibfield  {journal}
  {\bibinfo  {journal} {Phys. Rev. B}\ }\textbf {\bibinfo {volume} {80}},\
  \bibinfo {pages} {165126} (\bibinfo {year} {2009})}\BibitemShut {NoStop}%
\bibitem [{\citenamefont {Sakai}\ \emph {et~al.}(2009)\citenamefont {Sakai},
  \citenamefont {Motome},\ and\ \citenamefont {Imada}}]{sakai:2009}%
  \BibitemOpen
  \bibfield  {author} {\bibinfo {author} {\bibfnamefont {S.}~\bibnamefont
  {Sakai}}, \bibinfo {author} {\bibfnamefont {Y.}~\bibnamefont {Motome}}, \
  and\ \bibinfo {author} {\bibfnamefont {M.}~\bibnamefont {Imada}},\ }\href
  {\doibase 10.1103/PhysRevLett.102.056404} {\bibfield  {journal} {\bibinfo
  {journal} {Phys. Rev. Lett.}\ }\textbf {\bibinfo {volume} {102}},\ \bibinfo
  {pages} {056404} (\bibinfo {year} {2009})}\BibitemShut {NoStop}%
\bibitem [{\citenamefont {Lin}\ \emph {et~al.}(2010)\citenamefont {Lin},
  \citenamefont {Gull},\ and\ \citenamefont {Millis}}]{Lin10}%
  \BibitemOpen
  \bibfield  {author} {\bibinfo {author} {\bibfnamefont {N.}~\bibnamefont
  {Lin}}, \bibinfo {author} {\bibfnamefont {E.}~\bibnamefont {Gull}}, \ and\
  \bibinfo {author} {\bibfnamefont {A.~J.}\ \bibnamefont {Millis}},\ }\href
  {\doibase 10.1103/PhysRevB.82.045104} {\bibfield  {journal} {\bibinfo
  {journal} {Phys. Rev. B}\ }\textbf {\bibinfo {volume} {82}},\ \bibinfo
  {pages} {045104} (\bibinfo {year} {2010})}\BibitemShut {NoStop}%
\bibitem [{\citenamefont {Gull}\ \emph {et~al.}(2010)\citenamefont {Gull},
  \citenamefont {Ferrero}, \citenamefont {Parcollet}, \citenamefont {Georges},\
  and\ \citenamefont {Millis}}]{Gull10_clustercompare}%
  \BibitemOpen
  \bibfield  {author} {\bibinfo {author} {\bibfnamefont {E.}~\bibnamefont
  {Gull}}, \bibinfo {author} {\bibfnamefont {M.}~\bibnamefont {Ferrero}},
  \bibinfo {author} {\bibfnamefont {O.}~\bibnamefont {Parcollet}}, \bibinfo
  {author} {\bibfnamefont {A.}~\bibnamefont {Georges}}, \ and\ \bibinfo
  {author} {\bibfnamefont {A.~J.}\ \bibnamefont {Millis}},\ }\href {\doibase
  10.1103/PhysRevB.82.155101} {\bibfield  {journal} {\bibinfo  {journal} {Phys.
  Rev. B}\ }\textbf {\bibinfo {volume} {82}},\ \bibinfo {pages} {155101}
  (\bibinfo {year} {2010})}\BibitemShut {NoStop}%
\bibitem [{\citenamefont {Sordi}\ \emph {et~al.}(2012)\citenamefont {Sordi},
  \citenamefont {S\'emon}, \citenamefont {Haule},\ and\ \citenamefont
  {Tremblay}}]{Sordi12}%
  \BibitemOpen
  \bibfield  {author} {\bibinfo {author} {\bibfnamefont {G.}~\bibnamefont
  {Sordi}}, \bibinfo {author} {\bibfnamefont {P.}~\bibnamefont {S\'emon}},
  \bibinfo {author} {\bibfnamefont {K.}~\bibnamefont {Haule}}, \ and\ \bibinfo
  {author} {\bibfnamefont {A.-M.~S.}\ \bibnamefont {Tremblay}},\ }\href
  {\doibase 10.1103/PhysRevLett.108.216401} {\bibfield  {journal} {\bibinfo
  {journal} {Phys. Rev. Lett.}\ }\textbf {\bibinfo {volume} {108}},\ \bibinfo
  {pages} {216401} (\bibinfo {year} {2012})}\BibitemShut {NoStop}%
\bibitem [{\citenamefont {Lin}\ \emph {et~al.}(2012)\citenamefont {Lin},
  \citenamefont {Gull},\ and\ \citenamefont {Millis}}]{Lin12}%
  \BibitemOpen
  \bibfield  {author} {\bibinfo {author} {\bibfnamefont {N.}~\bibnamefont
  {Lin}}, \bibinfo {author} {\bibfnamefont {E.}~\bibnamefont {Gull}}, \ and\
  \bibinfo {author} {\bibfnamefont {A.~J.}\ \bibnamefont {Millis}},\ }\href
  {\doibase 10.1103/PhysRevLett.109.106401} {\bibfield  {journal} {\bibinfo
  {journal} {Phys. Rev. Lett.}\ }\textbf {\bibinfo {volume} {109}},\ \bibinfo
  {pages} {106401} (\bibinfo {year} {2012})}\BibitemShut {NoStop}%
\bibitem [{\citenamefont {Merino}\ and\ \citenamefont
  {Gunnarsson}(2014)}]{Merino14}%
  \BibitemOpen
  \bibfield  {author} {\bibinfo {author} {\bibfnamefont {J.}~\bibnamefont
  {Merino}}\ and\ \bibinfo {author} {\bibfnamefont {O.}~\bibnamefont
  {Gunnarsson}},\ }\href {\doibase 10.1103/PhysRevB.89.245130} {\bibfield
  {journal} {\bibinfo  {journal} {Phys. Rev. B}\ }\textbf {\bibinfo {volume}
  {89}},\ \bibinfo {pages} {245130} (\bibinfo {year} {2014})}\BibitemShut
  {NoStop}%
\bibitem [{\citenamefont {Gunnarsson}\ \emph
  {et~al.}(2015{\natexlab{a}})\citenamefont {Gunnarsson}, \citenamefont
  {Sch\"afer}, \citenamefont {LeBlanc}, \citenamefont {Gull}, \citenamefont
  {Merino}, \citenamefont {Sangiovanni}, \citenamefont {Rohringer},\ and\
  \citenamefont {Toschi}}]{Gunnarsson15}%
  \BibitemOpen
  \bibfield  {author} {\bibinfo {author} {\bibfnamefont {O.}~\bibnamefont
  {Gunnarsson}}, \bibinfo {author} {\bibfnamefont {T.}~\bibnamefont
  {Sch\"afer}}, \bibinfo {author} {\bibfnamefont {J.~P.~F.}\ \bibnamefont
  {LeBlanc}}, \bibinfo {author} {\bibfnamefont {E.}~\bibnamefont {Gull}},
  \bibinfo {author} {\bibfnamefont {J.}~\bibnamefont {Merino}}, \bibinfo
  {author} {\bibfnamefont {G.}~\bibnamefont {Sangiovanni}}, \bibinfo {author}
  {\bibfnamefont {G.}~\bibnamefont {Rohringer}}, \ and\ \bibinfo {author}
  {\bibfnamefont {A.}~\bibnamefont {Toschi}},\ }\href {\doibase
  10.1103/PhysRevLett.114.236402} {\bibfield  {journal} {\bibinfo  {journal}
  {Phys. Rev. Lett.}\ }\textbf {\bibinfo {volume} {114}},\ \bibinfo {pages}
  {236402} (\bibinfo {year} {2015}{\natexlab{a}})}\BibitemShut {NoStop}%
\bibitem [{\citenamefont {Chen}\ \emph {et~al.}(2017)\citenamefont {Chen},
  \citenamefont {LeBlanc},\ and\ \citenamefont {Gull}}]{Chen17}%
  \BibitemOpen
  \bibfield  {author} {\bibinfo {author} {\bibfnamefont {X.}~\bibnamefont
  {Chen}}, \bibinfo {author} {\bibfnamefont {J.~P.~F.}\ \bibnamefont
  {LeBlanc}}, \ and\ \bibinfo {author} {\bibfnamefont {E.}~\bibnamefont
  {Gull}},\ }\href {http://dx.doi.org/10.1038/ncomms14986} {\bibfield
  {journal} {\bibinfo  {journal} {Nature Communications}\ }\textbf {\bibinfo
  {volume} {8}},\ \bibinfo {pages} {14986} (\bibinfo {year}
  {2017})}\BibitemShut {NoStop}%
\bibitem [{\citenamefont {Lichtenstein}\ and\ \citenamefont
  {Katsnelson}(2000)}]{Lichtenstein00}%
  \BibitemOpen
  \bibfield  {author} {\bibinfo {author} {\bibfnamefont {A.~I.}\ \bibnamefont
  {Lichtenstein}}\ and\ \bibinfo {author} {\bibfnamefont {M.~I.}\ \bibnamefont
  {Katsnelson}},\ }\href {\doibase 10.1103/PhysRevB.62.R9283} {\bibfield
  {journal} {\bibinfo  {journal} {Phys. Rev. B}\ }\textbf {\bibinfo {volume}
  {62}},\ \bibinfo {pages} {R9283} (\bibinfo {year} {2000})}\BibitemShut
  {NoStop}%
\bibitem [{\citenamefont {Maier}\ \emph {et~al.}(2008)\citenamefont {Maier},
  \citenamefont {Poilblanc},\ and\ \citenamefont {Scalapino}}]{Maier08}%
  \BibitemOpen
  \bibfield  {author} {\bibinfo {author} {\bibfnamefont {T.~A.}\ \bibnamefont
  {Maier}}, \bibinfo {author} {\bibfnamefont {D.}~\bibnamefont {Poilblanc}}, \
  and\ \bibinfo {author} {\bibfnamefont {D.~J.}\ \bibnamefont {Scalapino}},\
  }\href {\doibase 10.1103/PhysRevLett.100.237001} {\bibfield  {journal}
  {\bibinfo  {journal} {Phys. Rev. Lett.}\ }\textbf {\bibinfo {volume} {100}},\
  \bibinfo {pages} {237001} (\bibinfo {year} {2008})}\BibitemShut {NoStop}%
\bibitem [{\citenamefont {Civelli}(2009)}]{Civelli09}%
  \BibitemOpen
  \bibfield  {author} {\bibinfo {author} {\bibfnamefont {M.}~\bibnamefont
  {Civelli}},\ }\href {\doibase 10.1103/PhysRevB.79.195113} {\bibfield
  {journal} {\bibinfo  {journal} {Phys. Rev. B}\ }\textbf {\bibinfo {volume}
  {79}},\ \bibinfo {pages} {195113} (\bibinfo {year} {2009})}\BibitemShut
  {NoStop}%
\bibitem [{\citenamefont {Gull}\ and\ \citenamefont
  {Millis}(2012)}]{Gull12_energy}%
  \BibitemOpen
  \bibfield  {author} {\bibinfo {author} {\bibfnamefont {E.}~\bibnamefont
  {Gull}}\ and\ \bibinfo {author} {\bibfnamefont {A.~J.}\ \bibnamefont
  {Millis}},\ }\href {\doibase 10.1103/PhysRevB.86.241106} {\bibfield
  {journal} {\bibinfo  {journal} {Phys. Rev. B}\ }\textbf {\bibinfo {volume}
  {86}},\ \bibinfo {pages} {241106} (\bibinfo {year} {2012})}\BibitemShut
  {NoStop}%
\bibitem [{\citenamefont {Gull}\ \emph {et~al.}(2013)\citenamefont {Gull},
  \citenamefont {Parcollet},\ and\ \citenamefont {Millis}}]{Gull13_super}%
  \BibitemOpen
  \bibfield  {author} {\bibinfo {author} {\bibfnamefont {E.}~\bibnamefont
  {Gull}}, \bibinfo {author} {\bibfnamefont {O.}~\bibnamefont {Parcollet}}, \
  and\ \bibinfo {author} {\bibfnamefont {A.~J.}\ \bibnamefont {Millis}},\
  }\href {\doibase 10.1103/PhysRevLett.110.216405} {\bibfield  {journal}
  {\bibinfo  {journal} {Phys. Rev. Lett.}\ }\textbf {\bibinfo {volume} {110}},\
  \bibinfo {pages} {216405} (\bibinfo {year} {2013})}\BibitemShut {NoStop}%
\bibitem [{\citenamefont {Gull}\ and\ \citenamefont
  {Millis}(2013)}]{Gull13_raman}%
  \BibitemOpen
  \bibfield  {author} {\bibinfo {author} {\bibfnamefont {E.}~\bibnamefont
  {Gull}}\ and\ \bibinfo {author} {\bibfnamefont {A.~J.}\ \bibnamefont
  {Millis}},\ }\href {\doibase 10.1103/PhysRevB.88.075127} {\bibfield
  {journal} {\bibinfo  {journal} {Phys. Rev. B}\ }\textbf {\bibinfo {volume}
  {88}},\ \bibinfo {pages} {075127} (\bibinfo {year} {2013})}\BibitemShut
  {NoStop}%
\bibitem [{\citenamefont {Sordi}\ \emph {et~al.}(2013)\citenamefont {Sordi},
  \citenamefont {S\'emon}, \citenamefont {Haule},\ and\ \citenamefont
  {Tremblay}}]{Sordi13}%
  \BibitemOpen
  \bibfield  {author} {\bibinfo {author} {\bibfnamefont {G.}~\bibnamefont
  {Sordi}}, \bibinfo {author} {\bibfnamefont {P.}~\bibnamefont {S\'emon}},
  \bibinfo {author} {\bibfnamefont {K.}~\bibnamefont {Haule}}, \ and\ \bibinfo
  {author} {\bibfnamefont {A.-M.~S.}\ \bibnamefont {Tremblay}},\ }\href
  {\doibase 10.1103/PhysRevB.87.041101} {\bibfield  {journal} {\bibinfo
  {journal} {Phys. Rev. B}\ }\textbf {\bibinfo {volume} {87}},\ \bibinfo
  {pages} {041101} (\bibinfo {year} {2013})}\BibitemShut {NoStop}%
\bibitem [{\citenamefont {Gull}\ and\ \citenamefont
  {Millis}(2014)}]{Gull14_pairing}%
  \BibitemOpen
  \bibfield  {author} {\bibinfo {author} {\bibfnamefont {E.}~\bibnamefont
  {Gull}}\ and\ \bibinfo {author} {\bibfnamefont {A.~J.}\ \bibnamefont
  {Millis}},\ }\href {\doibase 10.1103/PhysRevB.90.041110} {\bibfield
  {journal} {\bibinfo  {journal} {Phys. Rev. B}\ }\textbf {\bibinfo {volume}
  {90}},\ \bibinfo {pages} {041110} (\bibinfo {year} {2014})}\BibitemShut
  {NoStop}%
\bibitem [{\citenamefont {Gull}\ and\ \citenamefont
  {Millis}(2015)}]{Gull15_qp}%
  \BibitemOpen
  \bibfield  {author} {\bibinfo {author} {\bibfnamefont {E.}~\bibnamefont
  {Gull}}\ and\ \bibinfo {author} {\bibfnamefont {A.~J.}\ \bibnamefont
  {Millis}},\ }\href {\doibase 10.1103/PhysRevB.91.085116} {\bibfield
  {journal} {\bibinfo  {journal} {Phys. Rev. B}\ }\textbf {\bibinfo {volume}
  {91}},\ \bibinfo {pages} {085116} (\bibinfo {year} {2015})}\BibitemShut
  {NoStop}%
\bibitem [{\citenamefont {Chen}\ \emph {et~al.}(2015)\citenamefont {Chen},
  \citenamefont {LeBlanc},\ and\ \citenamefont {Gull}}]{Chen15}%
  \BibitemOpen
  \bibfield  {author} {\bibinfo {author} {\bibfnamefont {X.}~\bibnamefont
  {Chen}}, \bibinfo {author} {\bibfnamefont {J.~P.~F.}\ \bibnamefont
  {LeBlanc}}, \ and\ \bibinfo {author} {\bibfnamefont {E.}~\bibnamefont
  {Gull}},\ }\href {\doibase 10.1103/PhysRevLett.115.116402} {\bibfield
  {journal} {\bibinfo  {journal} {Phys. Rev. Lett.}\ }\textbf {\bibinfo
  {volume} {115}},\ \bibinfo {pages} {116402} (\bibinfo {year}
  {2015})}\BibitemShut {NoStop}%
\bibitem [{\citenamefont {{Maier}}\ \emph {et~al.}(2015)\citenamefont
  {{Maier}}, \citenamefont {{Staar}},\ and\ \citenamefont
  {{Scalapino}}}]{Maier15}%
  \BibitemOpen
  \bibfield  {author} {\bibinfo {author} {\bibfnamefont {T.~A.}\ \bibnamefont
  {{Maier}}}, \bibinfo {author} {\bibfnamefont {P.}~\bibnamefont {{Staar}}}, \
  and\ \bibinfo {author} {\bibfnamefont {D.~J.}\ \bibnamefont {{Scalapino}}},\
  }\href@noop {} {\bibfield  {journal} {\bibinfo  {journal} {arXiv e-prints}\
  ,\ \bibinfo {eid} {arXiv:1507.06206}} (\bibinfo {year} {2015})},\ \Eprint
  {http://arxiv.org/abs/1507.06206} {arXiv:1507.06206 [cond-mat.supr-con]}
  \BibitemShut {NoStop}%
\bibitem [{\citenamefont {{Maier}}\ and\ \citenamefont
  {{Scalapino}}(2018)}]{Maier18}%
  \BibitemOpen
  \bibfield  {author} {\bibinfo {author} {\bibfnamefont {T.~A.}\ \bibnamefont
  {{Maier}}}\ and\ \bibinfo {author} {\bibfnamefont {D.~J.}\ \bibnamefont
  {{Scalapino}}},\ }\href@noop {} {\bibfield  {journal} {\bibinfo  {journal}
  {arXiv e-prints}\ ,\ \bibinfo {eid} {arXiv:1810.10043}} (\bibinfo {year}
  {2018})},\ \Eprint {http://arxiv.org/abs/1810.10043} {arXiv:1810.10043
  [cond-mat.supr-con]} \BibitemShut {NoStop}%
\bibitem [{\citenamefont {Gunnarsson}\ \emph
  {et~al.}(2015{\natexlab{b}})\citenamefont {Gunnarsson}, \citenamefont
  {Sch\"afer}, \citenamefont {LeBlanc}, \citenamefont {Gull}, \citenamefont
  {Merino}, \citenamefont {Sangiovanni}, \citenamefont {Rohringer},\ and\
  \citenamefont {Toschi}}]{gunnarsson:2015}%
  \BibitemOpen
  \bibfield  {author} {\bibinfo {author} {\bibfnamefont {O.}~\bibnamefont
  {Gunnarsson}}, \bibinfo {author} {\bibfnamefont {T.}~\bibnamefont
  {Sch\"afer}}, \bibinfo {author} {\bibfnamefont {J.~P.~F.}\ \bibnamefont
  {LeBlanc}}, \bibinfo {author} {\bibfnamefont {E.}~\bibnamefont {Gull}},
  \bibinfo {author} {\bibfnamefont {J.}~\bibnamefont {Merino}}, \bibinfo
  {author} {\bibfnamefont {G.}~\bibnamefont {Sangiovanni}}, \bibinfo {author}
  {\bibfnamefont {G.}~\bibnamefont {Rohringer}}, \ and\ \bibinfo {author}
  {\bibfnamefont {A.}~\bibnamefont {Toschi}},\ }\href {\doibase
  10.1103/PhysRevLett.114.236402} {\bibfield  {journal} {\bibinfo  {journal}
  {Phys. Rev. Lett.}\ }\textbf {\bibinfo {volume} {114}},\ \bibinfo {pages}
  {236402} (\bibinfo {year} {2015}{\natexlab{b}})}\BibitemShut {NoStop}%
\bibitem [{\citenamefont {Delannoy}\ \emph {et~al.}(2009)\citenamefont
  {Delannoy}, \citenamefont {Gingras}, \citenamefont {Holdsworth},\ and\
  \citenamefont {Tremblay}}]{delannoy:2009}%
  \BibitemOpen
  \bibfield  {author} {\bibinfo {author} {\bibfnamefont {J.-Y.~P.}\
  \bibnamefont {Delannoy}}, \bibinfo {author} {\bibfnamefont {M.~J.~P.}\
  \bibnamefont {Gingras}}, \bibinfo {author} {\bibfnamefont {P.~C.~W.}\
  \bibnamefont {Holdsworth}}, \ and\ \bibinfo {author} {\bibfnamefont
  {A.-M.~S.}\ \bibnamefont {Tremblay}},\ }\href {\doibase
  10.1103/PhysRevB.79.235130} {\bibfield  {journal} {\bibinfo  {journal} {Phys.
  Rev. B}\ }\textbf {\bibinfo {volume} {79}},\ \bibinfo {pages} {235130}
  (\bibinfo {year} {2009})}\BibitemShut {NoStop}%
\bibitem [{\citenamefont {Jia}\ \emph {et~al.}(2014)\citenamefont {Jia},
  \citenamefont {Nowadnick}, \citenamefont {Wohlfeld}, \citenamefont {Kung},
  \citenamefont {Chen}, \citenamefont {Johnston}, \citenamefont {Tohyama},
  \citenamefont {Moritz},\ and\ \citenamefont {Devereaux}}]{Jia14}%
  \BibitemOpen
  \bibfield  {author} {\bibinfo {author} {\bibfnamefont {C.~J.}\ \bibnamefont
  {Jia}}, \bibinfo {author} {\bibfnamefont {E.~A.}\ \bibnamefont {Nowadnick}},
  \bibinfo {author} {\bibfnamefont {K.}~\bibnamefont {Wohlfeld}}, \bibinfo
  {author} {\bibfnamefont {Y.~F.}\ \bibnamefont {Kung}}, \bibinfo {author}
  {\bibfnamefont {C.~C.}\ \bibnamefont {Chen}}, \bibinfo {author}
  {\bibfnamefont {S.}~\bibnamefont {Johnston}}, \bibinfo {author}
  {\bibfnamefont {T.}~\bibnamefont {Tohyama}}, \bibinfo {author} {\bibfnamefont
  {B.}~\bibnamefont {Moritz}}, \ and\ \bibinfo {author} {\bibfnamefont {T.~P.}\
  \bibnamefont {Devereaux}},\ }\href {http://dx.doi.org/10.1038/ncomms4314}
  {\bibfield  {journal} {\bibinfo  {journal} {Nature Communications}\ }\textbf
  {\bibinfo {volume} {5}},\ \bibinfo {pages} {3314 EP } (\bibinfo {year}
  {2014})}\BibitemShut {NoStop}%
\bibitem [{\citenamefont {Zheng}\ \emph {et~al.}(2017)\citenamefont {Zheng},
  \citenamefont {Chung}, \citenamefont {Corboz}, \citenamefont {Ehlers},
  \citenamefont {Qin}, \citenamefont {Noack}, \citenamefont {Shi},
  \citenamefont {White}, \citenamefont {Zhang},\ and\ \citenamefont
  {Chan}}]{Zheng17}%
  \BibitemOpen
  \bibfield  {author} {\bibinfo {author} {\bibfnamefont {B.-X.}\ \bibnamefont
  {Zheng}}, \bibinfo {author} {\bibfnamefont {C.-M.}\ \bibnamefont {Chung}},
  \bibinfo {author} {\bibfnamefont {P.}~\bibnamefont {Corboz}}, \bibinfo
  {author} {\bibfnamefont {G.}~\bibnamefont {Ehlers}}, \bibinfo {author}
  {\bibfnamefont {M.-P.}\ \bibnamefont {Qin}}, \bibinfo {author} {\bibfnamefont
  {R.~M.}\ \bibnamefont {Noack}}, \bibinfo {author} {\bibfnamefont
  {H.}~\bibnamefont {Shi}}, \bibinfo {author} {\bibfnamefont {S.~R.}\
  \bibnamefont {White}}, \bibinfo {author} {\bibfnamefont {S.}~\bibnamefont
  {Zhang}}, \ and\ \bibinfo {author} {\bibfnamefont {G.~K.-L.}\ \bibnamefont
  {Chan}},\ }\href {\doibase 10.1126/science.aam7127} {\bibfield  {journal}
  {\bibinfo  {journal} {Science}\ }\textbf {\bibinfo {volume} {358}},\ \bibinfo
  {pages} {1155} (\bibinfo {year} {2017})},\ \Eprint
  {http://arxiv.org/abs/http://science.sciencemag.org/content/358/6367/1155.full.pdf}
  {http://science.sciencemag.org/content/358/6367/1155.full.pdf} \BibitemShut
  {NoStop}%
\bibitem [{\citenamefont {Huang}\ \emph {et~al.}(2017)\citenamefont {Huang},
  \citenamefont {Scalapino}, \citenamefont {Maier}, \citenamefont {Moritz},\
  and\ \citenamefont {Devereaux}}]{Huang:2017}%
  \BibitemOpen
  \bibfield  {author} {\bibinfo {author} {\bibfnamefont {E.~W.}\ \bibnamefont
  {Huang}}, \bibinfo {author} {\bibfnamefont {D.~J.}\ \bibnamefont
  {Scalapino}}, \bibinfo {author} {\bibfnamefont {T.~A.}\ \bibnamefont
  {Maier}}, \bibinfo {author} {\bibfnamefont {B.}~\bibnamefont {Moritz}}, \
  and\ \bibinfo {author} {\bibfnamefont {T.~P.}\ \bibnamefont {Devereaux}},\
  }\href {\doibase 10.1103/PhysRevB.96.020503} {\bibfield  {journal} {\bibinfo
  {journal} {Phys. Rev. B}\ }\textbf {\bibinfo {volume} {96}},\ \bibinfo
  {pages} {020503} (\bibinfo {year} {2017})}\BibitemShut {NoStop}%
\bibitem [{\citenamefont {Rubtsov}\ \emph {et~al.}(2008)\citenamefont
  {Rubtsov}, \citenamefont {Katsnelson},\ and\ \citenamefont
  {Lichtenstein}}]{Rubtsov08}%
  \BibitemOpen
  \bibfield  {author} {\bibinfo {author} {\bibfnamefont {A.~N.}\ \bibnamefont
  {Rubtsov}}, \bibinfo {author} {\bibfnamefont {M.~I.}\ \bibnamefont
  {Katsnelson}}, \ and\ \bibinfo {author} {\bibfnamefont {A.~I.}\ \bibnamefont
  {Lichtenstein}},\ }\href {\doibase 10.1103/PhysRevB.77.033101} {\bibfield
  {journal} {\bibinfo  {journal} {Phys. Rev. B}\ }\textbf {\bibinfo {volume}
  {77}},\ \bibinfo {pages} {033101} (\bibinfo {year} {2008})}\BibitemShut
  {NoStop}%
\bibitem [{\citenamefont {Georges}\ \emph {et~al.}(1996)\citenamefont
  {Georges}, \citenamefont {Kotliar}, \citenamefont {Krauth},\ and\
  \citenamefont {Rozenberg}}]{Georges96}%
  \BibitemOpen
  \bibfield  {author} {\bibinfo {author} {\bibfnamefont {A.}~\bibnamefont
  {Georges}}, \bibinfo {author} {\bibfnamefont {G.}~\bibnamefont {Kotliar}},
  \bibinfo {author} {\bibfnamefont {W.}~\bibnamefont {Krauth}}, \ and\ \bibinfo
  {author} {\bibfnamefont {M.~J.}\ \bibnamefont {Rozenberg}},\ }\href {\doibase
  10.1103/RevModPhys.68.13} {\bibfield  {journal} {\bibinfo  {journal} {Rev.
  Mod. Phys.}\ }\textbf {\bibinfo {volume} {68}},\ \bibinfo {pages} {13}
  (\bibinfo {year} {1996})}\BibitemShut {NoStop}%
\bibitem [{\citenamefont {Iskakov}\ \emph {et~al.}(2016)\citenamefont
  {Iskakov}, \citenamefont {Antipov},\ and\ \citenamefont {Gull}}]{Iskakov16}%
  \BibitemOpen
  \bibfield  {author} {\bibinfo {author} {\bibfnamefont {S.}~\bibnamefont
  {Iskakov}}, \bibinfo {author} {\bibfnamefont {A.~E.}\ \bibnamefont
  {Antipov}}, \ and\ \bibinfo {author} {\bibfnamefont {E.}~\bibnamefont
  {Gull}},\ }\href {\doibase 10.1103/PhysRevB.94.035102} {\bibfield  {journal}
  {\bibinfo  {journal} {Phys. Rev. B}\ }\textbf {\bibinfo {volume} {94}},\
  \bibinfo {pages} {035102} (\bibinfo {year} {2016})}\BibitemShut {NoStop}%
\bibitem [{\citenamefont {Scalapino}(2007)}]{scalapino:2007}%
  \BibitemOpen
  \bibfield  {author} {\bibinfo {author} {\bibfnamefont {D.}~\bibnamefont
  {Scalapino}},\ }in\ \href@noop {} {\emph {\bibinfo {booktitle} {Handbook of
  High-Temperature Superconductivity}}},\ \bibinfo {editor} {edited by\
  \bibinfo {editor} {\bibfnamefont {J.}~\bibnamefont {Schrieffer}}\ and\
  \bibinfo {editor} {\bibfnamefont {J.}~\bibnamefont {Brooks}}}\ (\bibinfo
  {publisher} {Springer New York},\ \bibinfo {year} {2007})\ pp.\ \bibinfo
  {pages} {495--526}\BibitemShut {NoStop}%
\bibitem [{\citenamefont {LeBlanc}\ \emph {et~al.}(2015)\citenamefont
  {LeBlanc}, \citenamefont {Antipov}, \citenamefont {Becca}, \citenamefont
  {Bulik}, \citenamefont {Chan}, \citenamefont {Chung}, \citenamefont {Deng},
  \citenamefont {Ferrero}, \citenamefont {Henderson}, \citenamefont
  {Jim\'enez-Hoyos}, \citenamefont {Kozik}, \citenamefont {Liu}, \citenamefont
  {Millis}, \citenamefont {Prokof'ev}, \citenamefont {Qin}, \citenamefont
  {Scuseria}, \citenamefont {Shi}, \citenamefont {Svistunov}, \citenamefont
  {Tocchio}, \citenamefont {Tupitsyn}, \citenamefont {White}, \citenamefont
  {Zhang}, \citenamefont {Zheng}, \citenamefont {Zhu},\ and\ \citenamefont
  {Gull}}]{LeBlanc15}%
  \BibitemOpen
  \bibfield  {author} {\bibinfo {author} {\bibfnamefont {J.~P.~F.}\
  \bibnamefont {LeBlanc}}, \bibinfo {author} {\bibfnamefont {A.~E.}\
  \bibnamefont {Antipov}}, \bibinfo {author} {\bibfnamefont {F.}~\bibnamefont
  {Becca}}, \bibinfo {author} {\bibfnamefont {I.~W.}\ \bibnamefont {Bulik}},
  \bibinfo {author} {\bibfnamefont {G.~K.-L.}\ \bibnamefont {Chan}}, \bibinfo
  {author} {\bibfnamefont {C.-M.}\ \bibnamefont {Chung}}, \bibinfo {author}
  {\bibfnamefont {Y.}~\bibnamefont {Deng}}, \bibinfo {author} {\bibfnamefont
  {M.}~\bibnamefont {Ferrero}}, \bibinfo {author} {\bibfnamefont {T.~M.}\
  \bibnamefont {Henderson}}, \bibinfo {author} {\bibfnamefont {C.~A.}\
  \bibnamefont {Jim\'enez-Hoyos}}, \bibinfo {author} {\bibfnamefont
  {E.}~\bibnamefont {Kozik}}, \bibinfo {author} {\bibfnamefont {X.-W.}\
  \bibnamefont {Liu}}, \bibinfo {author} {\bibfnamefont {A.~J.}\ \bibnamefont
  {Millis}}, \bibinfo {author} {\bibfnamefont {N.~V.}\ \bibnamefont
  {Prokof'ev}}, \bibinfo {author} {\bibfnamefont {M.}~\bibnamefont {Qin}},
  \bibinfo {author} {\bibfnamefont {G.~E.}\ \bibnamefont {Scuseria}}, \bibinfo
  {author} {\bibfnamefont {H.}~\bibnamefont {Shi}}, \bibinfo {author}
  {\bibfnamefont {B.~V.}\ \bibnamefont {Svistunov}}, \bibinfo {author}
  {\bibfnamefont {L.~F.}\ \bibnamefont {Tocchio}}, \bibinfo {author}
  {\bibfnamefont {I.~S.}\ \bibnamefont {Tupitsyn}}, \bibinfo {author}
  {\bibfnamefont {S.~R.}\ \bibnamefont {White}}, \bibinfo {author}
  {\bibfnamefont {S.}~\bibnamefont {Zhang}}, \bibinfo {author} {\bibfnamefont
  {B.-X.}\ \bibnamefont {Zheng}}, \bibinfo {author} {\bibfnamefont
  {Z.}~\bibnamefont {Zhu}}, \ and\ \bibinfo {author} {\bibfnamefont
  {E.}~\bibnamefont {Gull}} (\bibinfo {collaboration} {Simons Collaboration on
  the Many-Electron Problem}),\ }\href {\doibase 10.1103/PhysRevX.5.041041}
  {\bibfield  {journal} {\bibinfo  {journal} {Phys. Rev. X}\ }\textbf {\bibinfo
  {volume} {5}},\ \bibinfo {pages} {041041} (\bibinfo {year}
  {2015})}\BibitemShut {NoStop}%
\bibitem [{\citenamefont {Antipov}\ \emph {et~al.}(2015)\citenamefont
  {Antipov}, \citenamefont {LeBlanc},\ and\ \citenamefont {Gull}}]{Antipov15}%
  \BibitemOpen
  \bibfield  {author} {\bibinfo {author} {\bibfnamefont {A.~E.}\ \bibnamefont
  {Antipov}}, \bibinfo {author} {\bibfnamefont {J.~P.}\ \bibnamefont
  {LeBlanc}}, \ and\ \bibinfo {author} {\bibfnamefont {E.}~\bibnamefont
  {Gull}},\ }\href {\doibase http://dx.doi.org/10.1016/j.phpro.2015.07.107}
  {\bibfield  {journal} {\bibinfo  {journal} {Physics Procedia}\ }\textbf
  {\bibinfo {volume} {68}},\ \bibinfo {pages} {43 } (\bibinfo {year} {2015})},\
  \bibinfo {note} {proceedings of the 28th Workshop on Computer Simulation
  Studies in Condensed Matter Physics (CSP2015)}\BibitemShut {NoStop}%
\bibitem [{\citenamefont {Hafermann}\ \emph {et~al.}(2009)\citenamefont
  {Hafermann}, \citenamefont {Li}, \citenamefont {Rubtsov}, \citenamefont
  {Katsnelson}, \citenamefont {Lichtenstein},\ and\ \citenamefont
  {Monien}}]{Hafermann09}%
  \BibitemOpen
  \bibfield  {author} {\bibinfo {author} {\bibfnamefont {H.}~\bibnamefont
  {Hafermann}}, \bibinfo {author} {\bibfnamefont {G.}~\bibnamefont {Li}},
  \bibinfo {author} {\bibfnamefont {A.~N.}\ \bibnamefont {Rubtsov}}, \bibinfo
  {author} {\bibfnamefont {M.~I.}\ \bibnamefont {Katsnelson}}, \bibinfo
  {author} {\bibfnamefont {A.~I.}\ \bibnamefont {Lichtenstein}}, \ and\
  \bibinfo {author} {\bibfnamefont {H.}~\bibnamefont {Monien}},\ }\href
  {\doibase 10.1103/PhysRevLett.102.206401} {\bibfield  {journal} {\bibinfo
  {journal} {Phys. Rev. Lett.}\ }\textbf {\bibinfo {volume} {102}},\ \bibinfo
  {pages} {206401} (\bibinfo {year} {2009})}\BibitemShut {NoStop}%
\bibitem [{\citenamefont {LeBlanc}\ and\ \citenamefont
  {Gull}(2013)}]{LeBlanc13}%
  \BibitemOpen
  \bibfield  {author} {\bibinfo {author} {\bibfnamefont {J.~P.~F.}\
  \bibnamefont {LeBlanc}}\ and\ \bibinfo {author} {\bibfnamefont
  {E.}~\bibnamefont {Gull}},\ }\href {\doibase 10.1103/PhysRevB.88.155108}
  {\bibfield  {journal} {\bibinfo  {journal} {Phys. Rev. B}\ }\textbf {\bibinfo
  {volume} {88}},\ \bibinfo {pages} {155108} (\bibinfo {year}
  {2013})}\BibitemShut {NoStop}%
\bibitem [{\citenamefont {Blankenbecler}\ \emph {et~al.}(1981)\citenamefont
  {Blankenbecler}, \citenamefont {Scalapino},\ and\ \citenamefont
  {Sugar}}]{BSS81}%
  \BibitemOpen
  \bibfield  {author} {\bibinfo {author} {\bibfnamefont {R.}~\bibnamefont
  {Blankenbecler}}, \bibinfo {author} {\bibfnamefont {D.~J.}\ \bibnamefont
  {Scalapino}}, \ and\ \bibinfo {author} {\bibfnamefont {R.~L.}\ \bibnamefont
  {Sugar}},\ }\href {\doibase 10.1103/PhysRevD.24.2278} {\bibfield  {journal}
  {\bibinfo  {journal} {Phys. Rev. D}\ }\textbf {\bibinfo {volume} {24}},\
  \bibinfo {pages} {2278} (\bibinfo {year} {1981})}\BibitemShut {NoStop}%
\bibitem [{\citenamefont {Gaenko}\ \emph {et~al.}(2017)\citenamefont {Gaenko},
  \citenamefont {Antipov}, \citenamefont {Carcassi}, \citenamefont {Chen},
  \citenamefont {Chen}, \citenamefont {Dong}, \citenamefont {Gamper},
  \citenamefont {Gukelberger}, \citenamefont {Igarashi}, \citenamefont
  {Iskakov}, \citenamefont {KÃ¶nz}, \citenamefont {LeBlanc}, \citenamefont
  {Levy}, \citenamefont {Ma}, \citenamefont {Paki}, \citenamefont {Shinaoka},
  \citenamefont {Todo}, \citenamefont {Troyer},\ and\ \citenamefont
  {Gull}}]{Gaenko17}%
  \BibitemOpen
  \bibfield  {author} {\bibinfo {author} {\bibfnamefont {A.}~\bibnamefont
  {Gaenko}}, \bibinfo {author} {\bibfnamefont {A.}~\bibnamefont {Antipov}},
  \bibinfo {author} {\bibfnamefont {G.}~\bibnamefont {Carcassi}}, \bibinfo
  {author} {\bibfnamefont {T.}~\bibnamefont {Chen}}, \bibinfo {author}
  {\bibfnamefont {X.}~\bibnamefont {Chen}}, \bibinfo {author} {\bibfnamefont
  {Q.}~\bibnamefont {Dong}}, \bibinfo {author} {\bibfnamefont {L.}~\bibnamefont
  {Gamper}}, \bibinfo {author} {\bibfnamefont {J.}~\bibnamefont {Gukelberger}},
  \bibinfo {author} {\bibfnamefont {R.}~\bibnamefont {Igarashi}}, \bibinfo
  {author} {\bibfnamefont {S.}~\bibnamefont {Iskakov}}, \bibinfo {author}
  {\bibfnamefont {M.}~\bibnamefont {KÃ¶nz}}, \bibinfo {author} {\bibfnamefont
  {J.}~\bibnamefont {LeBlanc}}, \bibinfo {author} {\bibfnamefont
  {R.}~\bibnamefont {Levy}}, \bibinfo {author} {\bibfnamefont {P.}~\bibnamefont
  {Ma}}, \bibinfo {author} {\bibfnamefont {J.}~\bibnamefont {Paki}}, \bibinfo
  {author} {\bibfnamefont {H.}~\bibnamefont {Shinaoka}}, \bibinfo {author}
  {\bibfnamefont {S.}~\bibnamefont {Todo}}, \bibinfo {author} {\bibfnamefont
  {M.}~\bibnamefont {Troyer}}, \ and\ \bibinfo {author} {\bibfnamefont
  {E.}~\bibnamefont {Gull}},\ }\href {\doibase
  https://doi.org/10.1016/j.cpc.2016.12.009} {\bibfield  {journal} {\bibinfo
  {journal} {Computer Physics Communications}\ }\textbf {\bibinfo {volume}
  {213}},\ \bibinfo {pages} {235 } (\bibinfo {year} {2017})}\BibitemShut
  {NoStop}%
\bibitem [{\citenamefont {{Levy}}\ \emph {et~al.}(2017)\citenamefont {{Levy}},
  \citenamefont {{LeBlanc}},\ and\ \citenamefont {{Gull}}}]{Levy2016}%
  \BibitemOpen
  \bibfield  {author} {\bibinfo {author} {\bibfnamefont {R.}~\bibnamefont
  {{Levy}}}, \bibinfo {author} {\bibfnamefont {J.~P.~F.}\ \bibnamefont
  {{LeBlanc}}}, \ and\ \bibinfo {author} {\bibfnamefont {E.}~\bibnamefont
  {{Gull}}},\ }\href {\doibase 10.1016/j.cpc.2017.01.018} {\bibfield  {journal}
  {\bibinfo  {journal} {Comp. Phys. Comm.}\ }\textbf {\bibinfo {volume}
  {215}},\ \bibinfo {pages} {149} (\bibinfo {year} {2017})}\BibitemShut
  {NoStop}%
\bibitem [{\citenamefont {Rubtsov}\ \emph {et~al.}(2012)\citenamefont
  {Rubtsov}, \citenamefont {Katsnelson},\ and\ \citenamefont
  {Lichtenstein}}]{rubtsov:2012}%
  \BibitemOpen
  \bibfield  {author} {\bibinfo {author} {\bibfnamefont {A.~N.}\ \bibnamefont
  {Rubtsov}}, \bibinfo {author} {\bibfnamefont {M.~I.}\ \bibnamefont
  {Katsnelson}}, \ and\ \bibinfo {author} {\bibfnamefont {A.~I.}\ \bibnamefont
  {Lichtenstein}},\ }\href@noop {} {\bibfield  {journal} {\bibinfo  {journal}
  {Annals of Physics}\ }\textbf {\bibinfo {volume} {327}},\ \bibinfo {pages}
  {1320} (\bibinfo {year} {2012})}\BibitemShut {NoStop}%
\bibitem [{\citenamefont {van Loon}\ \emph {et~al.}(2014)\citenamefont {van
  Loon}, \citenamefont {Lichtenstein}, \citenamefont {Katsnelson},
  \citenamefont {Parcollet},\ and\ \citenamefont {Hafermann}}]{vanloon:2014}%
  \BibitemOpen
  \bibfield  {author} {\bibinfo {author} {\bibfnamefont {E.~G. C.~P.}\
  \bibnamefont {van Loon}}, \bibinfo {author} {\bibfnamefont {A.~I.}\
  \bibnamefont {Lichtenstein}}, \bibinfo {author} {\bibfnamefont {M.~I.}\
  \bibnamefont {Katsnelson}}, \bibinfo {author} {\bibfnamefont
  {O.}~\bibnamefont {Parcollet}}, \ and\ \bibinfo {author} {\bibfnamefont
  {H.}~\bibnamefont {Hafermann}},\ }\href {\doibase 10.1103/PhysRevB.90.235135}
  {\bibfield  {journal} {\bibinfo  {journal} {Phys. Rev. B}\ }\textbf {\bibinfo
  {volume} {90}},\ \bibinfo {pages} {235135} (\bibinfo {year}
  {2014})}\BibitemShut {NoStop}%
\bibitem [{\citenamefont {Headings}\ \emph {et~al.}(2010)\citenamefont
  {Headings}, \citenamefont {Hayden}, \citenamefont {Coldea},\ and\
  \citenamefont {Perring}}]{Headings10}%
  \BibitemOpen
  \bibfield  {author} {\bibinfo {author} {\bibfnamefont {N.~S.}\ \bibnamefont
  {Headings}}, \bibinfo {author} {\bibfnamefont {S.~M.}\ \bibnamefont
  {Hayden}}, \bibinfo {author} {\bibfnamefont {R.}~\bibnamefont {Coldea}}, \
  and\ \bibinfo {author} {\bibfnamefont {T.~G.}\ \bibnamefont {Perring}},\
  }\href {\doibase 10.1103/PhysRevLett.105.247001} {\bibfield  {journal}
  {\bibinfo  {journal} {Phys. Rev. Lett.}\ }\textbf {\bibinfo {volume} {105}},\
  \bibinfo {pages} {247001} (\bibinfo {year} {2010})}\BibitemShut {NoStop}%
\bibitem [{\citenamefont {Andersen}\ \emph {et~al.}(1995)\citenamefont
  {Andersen}, \citenamefont {Liechtenstein}, \citenamefont {Jepsen},\ and\
  \citenamefont {Paulsen}}]{Andersen95}%
  \BibitemOpen
  \bibfield  {author} {\bibinfo {author} {\bibfnamefont {O.}~\bibnamefont
  {Andersen}}, \bibinfo {author} {\bibfnamefont {A.}~\bibnamefont
  {Liechtenstein}}, \bibinfo {author} {\bibfnamefont {O.}~\bibnamefont
  {Jepsen}}, \ and\ \bibinfo {author} {\bibfnamefont {F.}~\bibnamefont
  {Paulsen}},\ }\href {\doibase https://doi.org/10.1016/0022-3697(95)00269-3}
  {\bibfield  {journal} {\bibinfo  {journal} {Journal of Physics and Chemistry
  of Solids}\ }\textbf {\bibinfo {volume} {56}},\ \bibinfo {pages} {1573 }
  (\bibinfo {year} {1995})},\ \bibinfo {note} {proceedings of the Conference on
  Spectroscopies in Novel Superconductors}\BibitemShut {NoStop}%
\bibitem [{\citenamefont {Dean}\ \emph {et~al.}(2013)\citenamefont {Dean},
  \citenamefont {Dellea}, \citenamefont {Springell}, \citenamefont
  {Yakhou-Harris}, \citenamefont {Kummer}, \citenamefont {Brookes},
  \citenamefont {Liu}, \citenamefont {Sun}, \citenamefont {Strle},
  \citenamefont {Schmitt}, \citenamefont {Braicovich}, \citenamefont
  {Ghiringhelli}, \citenamefont {Bo{\v z}ovi{\'c}},\ and\ \citenamefont
  {Hill}}]{Dean13}%
  \BibitemOpen
  \bibfield  {author} {\bibinfo {author} {\bibfnamefont {M.~P.~M.}\
  \bibnamefont {Dean}}, \bibinfo {author} {\bibfnamefont {G.}~\bibnamefont
  {Dellea}}, \bibinfo {author} {\bibfnamefont {R.~S.}\ \bibnamefont
  {Springell}}, \bibinfo {author} {\bibfnamefont {F.}~\bibnamefont
  {Yakhou-Harris}}, \bibinfo {author} {\bibfnamefont {K.}~\bibnamefont
  {Kummer}}, \bibinfo {author} {\bibfnamefont {N.~B.}\ \bibnamefont {Brookes}},
  \bibinfo {author} {\bibfnamefont {X.}~\bibnamefont {Liu}}, \bibinfo {author}
  {\bibfnamefont {Y.-J.}\ \bibnamefont {Sun}}, \bibinfo {author} {\bibfnamefont
  {J.}~\bibnamefont {Strle}}, \bibinfo {author} {\bibfnamefont
  {T.}~\bibnamefont {Schmitt}}, \bibinfo {author} {\bibfnamefont
  {L.}~\bibnamefont {Braicovich}}, \bibinfo {author} {\bibfnamefont
  {G.}~\bibnamefont {Ghiringhelli}}, \bibinfo {author} {\bibfnamefont
  {I.}~\bibnamefont {Bo{\v z}ovi{\'c}}}, \ and\ \bibinfo {author}
  {\bibfnamefont {J.~P.}\ \bibnamefont {Hill}},\ }\href
  {http://dx.doi.org/10.1038/nmat3723} {\bibfield  {journal} {\bibinfo
  {journal} {Nat Mater}\ }\textbf {\bibinfo {volume} {12}},\ \bibinfo {pages}
  {1019} (\bibinfo {year} {2013})}\BibitemShut {NoStop}%
\bibitem [{\citenamefont {Chan}\ \emph {et~al.}(2016)\citenamefont {Chan},
  \citenamefont {Dorow}, \citenamefont {Mangin-Thro}, \citenamefont {Tang},
  \citenamefont {Ge}, \citenamefont {Veit}, \citenamefont {Yu}, \citenamefont
  {Zhao}, \citenamefont {Christianson}, \citenamefont {Park}, \citenamefont
  {Sidis}, \citenamefont {Steffens}, \citenamefont {Abernathy}, \citenamefont
  {Bourges},\ and\ \citenamefont {Greven}}]{Chan16}%
  \BibitemOpen
  \bibfield  {author} {\bibinfo {author} {\bibfnamefont {M.~K.}\ \bibnamefont
  {Chan}}, \bibinfo {author} {\bibfnamefont {C.~J.}\ \bibnamefont {Dorow}},
  \bibinfo {author} {\bibfnamefont {L.}~\bibnamefont {Mangin-Thro}}, \bibinfo
  {author} {\bibfnamefont {Y.}~\bibnamefont {Tang}}, \bibinfo {author}
  {\bibfnamefont {Y.}~\bibnamefont {Ge}}, \bibinfo {author} {\bibfnamefont
  {M.~J.}\ \bibnamefont {Veit}}, \bibinfo {author} {\bibfnamefont
  {G.}~\bibnamefont {Yu}}, \bibinfo {author} {\bibfnamefont {X.}~\bibnamefont
  {Zhao}}, \bibinfo {author} {\bibfnamefont {A.~D.}\ \bibnamefont
  {Christianson}}, \bibinfo {author} {\bibfnamefont {J.~T.}\ \bibnamefont
  {Park}}, \bibinfo {author} {\bibfnamefont {Y.}~\bibnamefont {Sidis}},
  \bibinfo {author} {\bibfnamefont {P.}~\bibnamefont {Steffens}}, \bibinfo
  {author} {\bibfnamefont {D.~L.}\ \bibnamefont {Abernathy}}, \bibinfo {author}
  {\bibfnamefont {P.}~\bibnamefont {Bourges}}, \ and\ \bibinfo {author}
  {\bibfnamefont {M.}~\bibnamefont {Greven}},\ }\href
  {http://dx.doi.org/10.1038/ncomms10819} {\bibfield  {journal} {\bibinfo
  {journal} {Nature Communications}\ }\textbf {\bibinfo {volume} {7}},\
  \bibinfo {pages} {10819 EP } (\bibinfo {year} {2016})}\BibitemShut {NoStop}%
\bibitem [{\citenamefont {Das}(2012)}]{Das12}%
  \BibitemOpen
  \bibfield  {author} {\bibinfo {author} {\bibfnamefont {T.}~\bibnamefont
  {Das}},\ }\href {\doibase 10.1103/PhysRevB.86.054518} {\bibfield  {journal}
  {\bibinfo  {journal} {Phys. Rev. B}\ }\textbf {\bibinfo {volume} {86}},\
  \bibinfo {pages} {054518} (\bibinfo {year} {2012})}\BibitemShut {NoStop}%
\bibitem [{\citenamefont {Vishik}\ \emph {et~al.}(2014)\citenamefont {Vishik},
  \citenamefont {Bari\ifmmode \check{s}\else \v{s}\fi{}i\ifmmode~\acute{c}\else
  \'{c}\fi{}}, \citenamefont {Chan}, \citenamefont {Li}, \citenamefont {Xia},
  \citenamefont {Yu}, \citenamefont {Zhao}, \citenamefont {Lee}, \citenamefont
  {Meevasana}, \citenamefont {Devereaux}, \citenamefont {Greven},\ and\
  \citenamefont {Shen}}]{vishik:2014}%
  \BibitemOpen
  \bibfield  {author} {\bibinfo {author} {\bibfnamefont {I.~M.}\ \bibnamefont
  {Vishik}}, \bibinfo {author} {\bibfnamefont {N.}~\bibnamefont {Bari\ifmmode
  \check{s}\else \v{s}\fi{}i\ifmmode~\acute{c}\else \'{c}\fi{}}}, \bibinfo
  {author} {\bibfnamefont {M.~K.}\ \bibnamefont {Chan}}, \bibinfo {author}
  {\bibfnamefont {Y.}~\bibnamefont {Li}}, \bibinfo {author} {\bibfnamefont
  {D.~D.}\ \bibnamefont {Xia}}, \bibinfo {author} {\bibfnamefont
  {G.}~\bibnamefont {Yu}}, \bibinfo {author} {\bibfnamefont {X.}~\bibnamefont
  {Zhao}}, \bibinfo {author} {\bibfnamefont {W.~S.}\ \bibnamefont {Lee}},
  \bibinfo {author} {\bibfnamefont {W.}~\bibnamefont {Meevasana}}, \bibinfo
  {author} {\bibfnamefont {T.~P.}\ \bibnamefont {Devereaux}}, \bibinfo {author}
  {\bibfnamefont {M.}~\bibnamefont {Greven}}, \ and\ \bibinfo {author}
  {\bibfnamefont {Z.-X.}\ \bibnamefont {Shen}},\ }\href {\doibase
  10.1103/PhysRevB.89.195141} {\bibfield  {journal} {\bibinfo  {journal} {Phys.
  Rev. B}\ }\textbf {\bibinfo {volume} {89}},\ \bibinfo {pages} {195141}
  (\bibinfo {year} {2014})}\BibitemShut {NoStop}%
\bibitem [{\citenamefont {Rohringer}\ and\ \citenamefont
  {Toschi}(2016)}]{rohringer:2016}%
  \BibitemOpen
  \bibfield  {author} {\bibinfo {author} {\bibfnamefont {G.}~\bibnamefont
  {Rohringer}}\ and\ \bibinfo {author} {\bibfnamefont {A.}~\bibnamefont
  {Toschi}},\ }\href {\doibase 10.1103/PhysRevB.94.125144} {\bibfield
  {journal} {\bibinfo  {journal} {Phys. Rev. B}\ }\textbf {\bibinfo {volume}
  {94}},\ \bibinfo {pages} {125144} (\bibinfo {year} {2016})}\BibitemShut
  {NoStop}%
\bibitem [{\citenamefont {Gukelberger}\ \emph {et~al.}(2017)\citenamefont
  {Gukelberger}, \citenamefont {Kozik},\ and\ \citenamefont
  {Hafermann}}]{gukelberger:2016}%
  \BibitemOpen
  \bibfield  {author} {\bibinfo {author} {\bibfnamefont {J.}~\bibnamefont
  {Gukelberger}}, \bibinfo {author} {\bibfnamefont {E.}~\bibnamefont {Kozik}},
  \ and\ \bibinfo {author} {\bibfnamefont {H.}~\bibnamefont {Hafermann}},\
  }\href {\doibase 10.1103/PhysRevB.96.035152} {\bibfield  {journal} {\bibinfo
  {journal} {Phys. Rev. B}\ }\textbf {\bibinfo {volume} {96}},\ \bibinfo
  {pages} {035152} (\bibinfo {year} {2017})}\BibitemShut {NoStop}%
\end{thebibliography}%
\end{document}